\documentclass[conference,compsoc]{IEEEtran}

\usepackage[final]{changes}
\usepackage{color}
\usepackage{graphicx}
\graphicspath{{./figs/}}
\usepackage[labelformat=simple]{subcaption}
\usepackage{xspace}
\usepackage{multirow}
\usepackage{makecell} %
\usepackage{enumitem} %
\usepackage[ruled,vlined]{algorithm2e}
\usepackage[nolist,nohyperlinks]{acronym}

\usepackage{ulem}
\newcommand{\etal}{et~al.\xspace}
\newcommand{\wifi}{Wi-Fi\xspace}
\newcommand{\vsix}{IPv6\xspace}

\newcommand{\eg}{e.g.\xspace}
\newcommand{\ie}{i.e.\xspace}

\newcommand{\parhead}[1]{\medskip \noindent \textbf{#1}\hskip .1in}

\acrodefplural{WPS}[WPSes]{Wi-Fi Positioning Systems}

\ifdefined\isFinalized
\newcommand{\NewCommentType}[3]{\expandafter\newcommand\csname #1\endcsname[1]{\begin{deletedEnv}#1\end{deletedEnv} }}
\else
\newcommand{\NewCommentType}[3]{\expandafter\newcommand\csname #1\endcsname[1]{{\color{#2}{#3: ##1}} }}
\fi

\usepackage{hyperref}
\usepackage[override]{cmtt} %

\begin{document}
\date{}

\newcommand{\name}{$\mathsf{SystemName}$\xspace} %
\newcommand{\Name}{$\mathsf{SystemName}$\xspace} %

\NewCommentType{dml}{magenta}{dml}

\title{Surveilling the Masses with Wi-Fi\replaced{-Based Positioning Systems}{ Geolocation Services}}

\author{

{\rm Erik Rye}\\
 University of Maryland
 \and
 {\rm Dave Levin}\\
 University of Maryland
}

\maketitle

\begin{abstract}
\wifi\added{-based} Positioning Systems \added{(WPSes)} are used by \deleted{all} modern mobile devices to
\replaced{learn}{trilaterate} their position \replaced{using nearby \wifi access points as
landmarks}{from nearby \wifi access points}.
In this work, we show that \replaced{Apple's 
WPS}{these systems} \replaced{can be abused to create a privacy threat on
a}{are a privacy threat on a} global scale.
\added{We present an attack that allows an unprivileged attacker to amass}
a worldwide snapshot of \wifi BSSID geolocations in only a matter of days.
\deleted{at no cost to themselves.}
\added{Our attack makes few assumptions, merely exploiting the fact that
there are relatively few dense regions of allocated MAC address space.}
Applying this technique over the course of a year, we learned the precise
locations of over 2 billion BSSIDs around the world. %

The privacy implications of such massive datasets become more stark when
taken longitudinally, allowing the attacker to track devices' movements.
While most \wifi access points do not move for long periods of time, many
devices---like compact travel routers---are
specifically designed to be mobile.

\added{We present several case studies that demonstrate the types of attacks on
privacy that Apple's WPS enables:}
We track devices moving in and out of war zones (specifically Ukraine and
Gaza), the effects of natural disasters (specifically the fires in Maui), and
the possibility of targeted individual tracking by proxy---all by remotely
geolocating wireless access points.

\replaced{We provide recommendations to 
WPS operators and
\wifi access point manufacturers to enhance the privacy of hundreds of millions of
users worldwide.}{Finally, we provide recommendations for device vendors and operators of \wifi
positioning systems.}
\added{Finally, we detail our efforts at responsibly disclosing this privacy
vulnerability, and outline some mitigations that Apple and \wifi access point
manufacturers have implemented both independently and as a result of our work.}
 \end{abstract}

\section{Introduction}
\label{sec:intro}

Mobile devices increasingly rely on frequent, precise geolocation, both
for location-based services (e.g., driving directions, advertising,
gaming, localized search) as well as for tracking one's devices in the
case of loss or theft~\cite{police-auctions-oakland} (e.g., Apple's
Find My service).
Due to its high power consumption, GPS is not a viable solution for
such frequent geolocation needs.
Instead, Apple and Google operate \emph{Wi-Fi-based Positioning
Systems} (WPSes), which allow mobile devices to query a server for
their location based on the Wi-Fi access points they see.

At a high level\footnote{We provide more details of Apple's WPS in
\S\ref{sec:background}.}:
mobile devices that have used GPS to ascertain their location
periodically report to the WPS the Wi-Fi access points' MAC
addresses---known as BSSIDs---that they observe, along with their GPS
coordinates.
The WPS stores the reported locations of the BSSIDs at a server.
Then, other mobile devices who are unable or unwilling to use GPS can
query the server, providing a set of BSSIDs it sees and receiving an
estimated geolocation in return.

As prior work has
noted~\cite{tippenhauer2009attacks,feng2014vulnerability}, popular
WPSes (especially Apple's and Google's) are publicly accessible, and
they do not require devices querying the database to prove they
actually see the BSSIDs they claim to see.
In other words, one can query for \emph{any} arbitrary MAC address and,
if it is in the WPS's database, then it will return its location.
This design lends itself to relatively obvious
\emph{individually-targeted} attacks.
For instance, if a victim of intimate partner violence were to move to
an undisclosed location, their ex-partner could periodically query the
WPS with the BSSID of the victim's Wi-Fi access point (or travel modem,
Wi-Fi enabled TV, etc.) until its location appears, thereby divulging
the victim's location.
While dangerous, individually-targeted attacks like this require an
attacker to have \emph{a priori} knowledge about their targets, thereby
limiting the potential threat.

In this paper, we show that an unprivileged, weak attacker can \replaced{take
advantage of}{use}
\replaced{Apple's}{a}
WPS to perform \emph{mass surveillance} of users' Wi-Fi access points
virtually anywhere in the world, without any \emph{a priori} knowledge.
We present novel ways to query \replaced{Apple's WPS}{WPS servers} that
allow us to learn about
devices worldwide, and to exhaustively track devices into and out of
target geographic regions.
We perform a systematic empirical evaluation of the data available from
WPSes, finding that they span hundreds of millions of devices, and that
they allow us to monitor the movement of Wi-Fi access points and other
devices.

To demonstrate the potential our attack has at using WPSes for
open-source intelligence (OSINT), we present several case studies,
including:

\parhead{Russia-Ukraine War}
First, we use Apple's WPS to analyze device movements into and out of
Ukraine and Russia, gaining insights into their ongoing war that, to
the best of our knowledge, have yet to be made public.
We find what appear to be personal devices being brought by military
personnel into war
zones, exposing pre-deployment sites and military positions.
Our results also show individuals who have left Ukraine to a wide range
of countries, validating public reports of where Ukrainian refugees
have resettled.

\parhead{Israel-Hamas War in Gaza}
Second, we use Apple's WPS to track movements out of and within Gaza, as
well as the disappearance of devices throughout the Gaza Strip.
This case study shows that it is possible to use WPS data to track
extensive outages and loss of devices.

\medskip

Making matters worse, users whose devices are being tracked never opted
in to \added{Apple's} WPS in the first place, \replaced{nor did they have a way
to opt out when we conducted this study}
{nor do they appear to have a way to opt out}.
Merely being within Wi-Fi range of an Apple \deleted{or Google} device can lead
to a device's location and movements being made widely and publicly
available.
Indeed, we identify devices from \replaced{over 10,000}{15,307} distinct vendors
in Apple's WPS.
\deleted{alone.}
\added{Fortunately, in response to our responsible disclosure
(\S\ref{sec:disclosure}), Apple now provides a way for users to opt
their devices out.}
We discuss \added{additional} potential mitigations in
\S\ref{sec:remediation}.

\parhead{Contributions}
To summarize, this paper makes the following contributions:

\begin{itemize}

\item We introduce the first known techniques for using Wi-Fi-based
Positioning Services to perform \emph{mass surveillance} of Wi-Fi
		access point locations, movements, and outages.
		(\S\ref{sec:threat})

\item We perform an extensive and systematic evaluation of the extent to which
an unprivileged attacker can perform mass surveillance using Apple's WPS, building from no
        \emph{a priori} knowledge a corpus
of 490,940,552 BSSIDs in virtually all locations 
around the world. (\S\ref{sec:results})

\item We present several case studies of the mass surveillance attack:
movements in and out of military locations (within Ukraine and Gaza) as
		well as the aftermath of a natural disaster (the fires in Maui,
		Hawaii). (\S\ref{sec:casestudies})

\item We present several possible ways that various stakeholders could
	mitigate these attacks. (\S\ref{sec:remediation})

\item \replaced{We disclosed our findings to Apple and Google---both of which
	operate their own WPS---as well as two of the most salient
		manufacturers in our case studies: SpaceX and GL.iNet. We
		report on this disclosure process and the remediations Apple
		and SpaceX have taken in response.}
		{We have disclosed our findings to Google and are
		in the process of disclosing them to Apple.}
        \added{(\S\ref{sec:disclosure})}

\end{itemize}

\section{Background}
\label{sec:background}

In this section we discuss 802.11 hardware identifiers, commonly known as
\ac{MAC} addresses, and \added{the operation of} \acp{WPS}, which are systems that assist in geolocation
by using 802.11 identifiers as landmarks.

\begin{figure*}[t]
    \begin{subfigure}[c]{0.48\textwidth}
        \centering
    \includegraphics[width=0.75\linewidth]{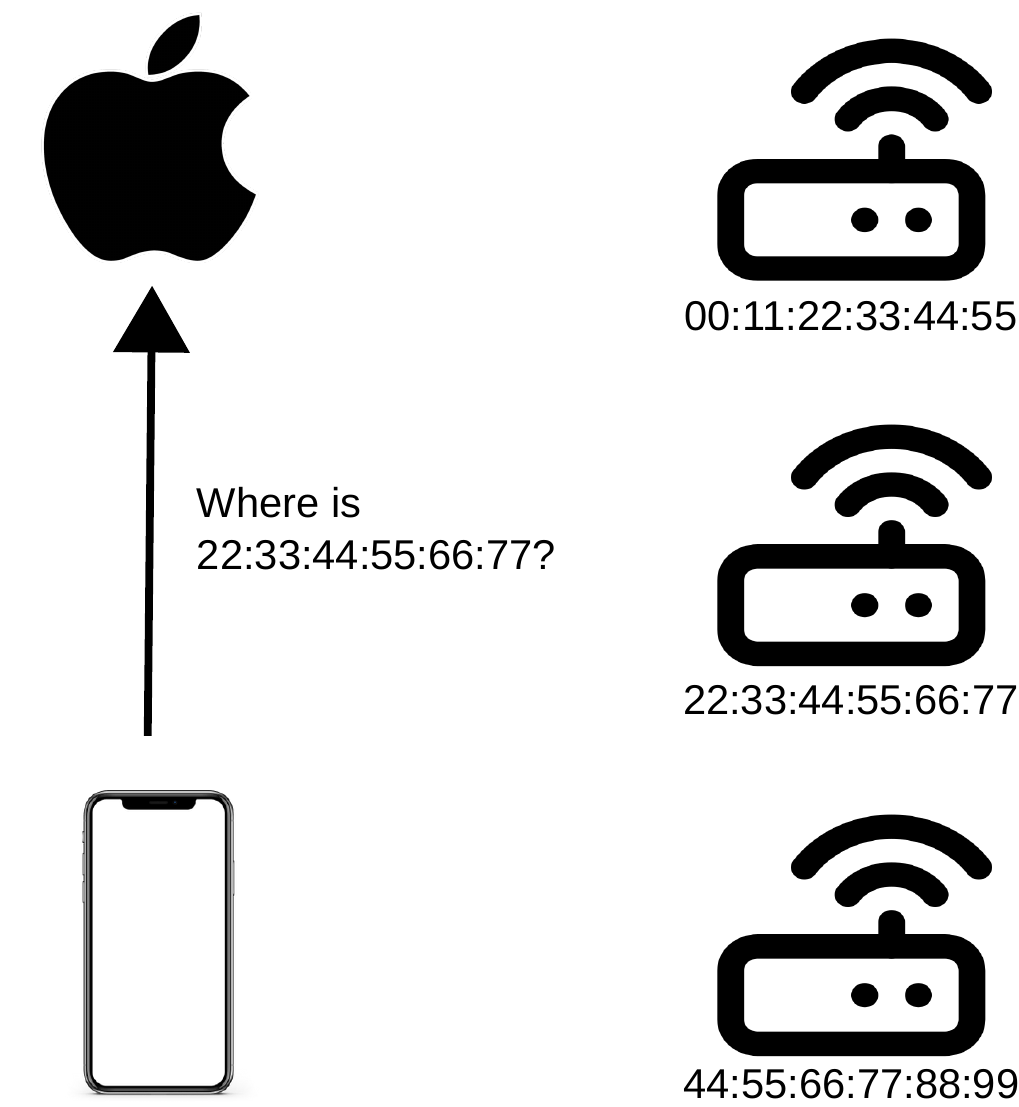}
        \caption{An Apple device queries the Apple \ac{WPS} for BSSIDs it
        detects in 802.11 scans to geolocate itself.}
    \label{fig:wpsquery}
  \end{subfigure}
    \hspace{1em}
    \begin{subfigure}[c]{0.48\textwidth}
        \centering
    \includegraphics[width=0.75\linewidth]{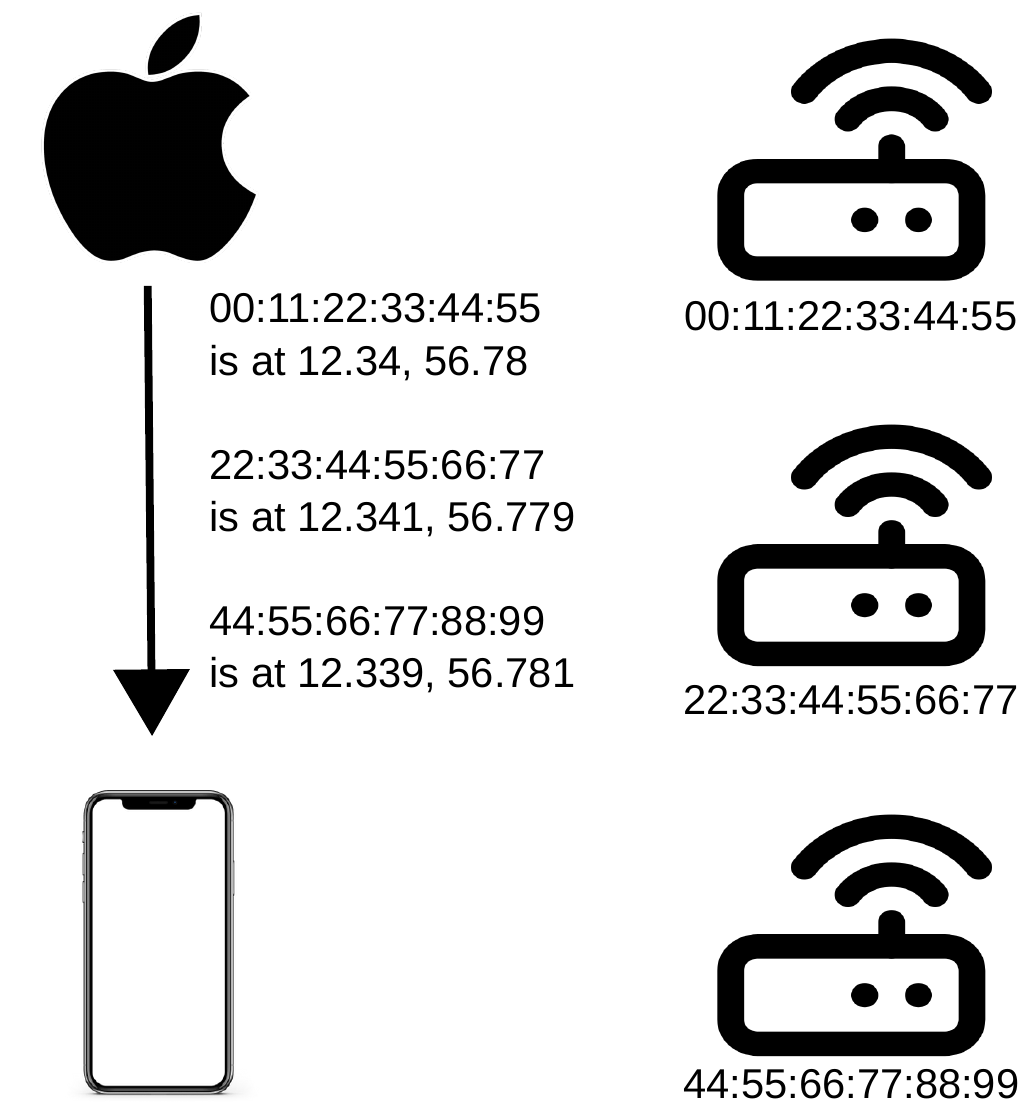}
        \caption{The Apple \ac{WPS} responds with the BSSID's geolocation as well as
        the geolocations of up to 400 additional nearby BSSIDs.}
        \label{fig:wpsresponse}
  \end{subfigure}
    \caption{An Apple device querying and receiving BSSID geolocations from the
    Apple \ac{WPS}. The \ac{WPS} is populated by other Apple devices that report
    their geolocations (derived from \eg GPS) and nearby BSSIDs, which the
    \ac{WPS} then uses as landmarks. \added{This figure shows 3 decimal digits of
    precision, though Apple routinely provides up to 8.}}
    \label{fig:wps}
\end{figure*}

\subsection{MAC Addresses and BSSIDs}
\label{subsec:macaddrs}

\ac{MAC} addresses are 48-bit link-layer identifiers used to identify a network
interface on a local network. MAC addresses are typically written as 12
hexadecimal digits, with each pair separated by colons or hyphens\added{, \eg,
\texttt{08:4A:93:2F:B1:07}}. When a \wifi
\ac{AP} creates an 802.11 network, the \ac{AP} and the devices connected to it
form a \emph{Basic Service Set (BSS)}. The MAC address the \wifi \ac{AP} uses
for that BSS is called a \ac{BSSID}. %
Note that there is a
many-to-one relationship between BSSIDs and \acp{AP}; one \ac{AP} can form
several 802.11 \wifi networks (e.g., an employee network and a guest
network) by using different BSSIDs for each.

Some bits of MAC addresses have semantic meaning. The lowest order bit
of the first byte, for example, indicates whether the MAC address is
unicast (when the bit is unset) or multicast (when set). \added{Of
particular note,} the second lowest order bit of the first byte---the
so-called Universal/Local (U/L) bit---indicates whether the MAC address
is globally assigned to a manufacturer by the IEEE (when the bit is
unset), or if the MAC address is locally assigned by the device.
Locally-assigned MAC addresses are often used for peer-to-peer
applications (which typically invert the U/L bit of the
globally-assigned MAC address) or by \wifi \acp{AP} advertising
multiple 802.11 \wifi networks by creating virtual interfaces using the
same physical hardware. \added{When devices choose random MAC
addresses, as most modern mobile 802.11 clients do for anti-tracking
protection, they also set the U/L bit.}

\subsection{\acfp{WPS}}

\acp{WPS}, which use BSSIDs as landmarks by which to geolocate
devices, are central to this work. \acp{WPS} are typically queried via an API.
These APIs are backed by
large-scale databases that contain the geolocations of hundreds of
millions of \wifi routers. 

\acp{WPS} obtain the locations of the BSSIDs in their databases through a
crowd-sourced network of sensors. In the case of Google and Apple, the devices
that populate their \acp{WPS} are mobile phones and other devices running their
operating systems. Periodically, these devices will forward their locations
(obtained through GPS or trilateration using cell towers as landmarks) along
with nearby \ac{AP} BSSIDs and their received signal strengths. If enough
sensors detect a BSSID \added{for a long enough period of time}, the BSSID and its estimated location are made
entries in the \ac{WPS}. 

These \acp{WPS} typically operate in one of two ways. First,
a client may submit a list of nearby \wifi \ac{AP} \acp{BSSID} and signal
strengths through the API. The positioning system then calculates the position
of the client that observed the \acp{AP}, and responds to the client with the
computed position. Google's \wifi Geolocation API~\cite{googlegeo} works in this
manner, requiring at least 2 \acp{BSSID} in order to calculate a client's
position. 

Conversely, other \wifi geolocation systems put the onus of
trilateration on the client. These APIs also accept a list of nearby \ac{AP}
\acp{BSSID}; instead of computing the client's location based off the set of
observed \acp{AP} and their received signal strengths, the API returns the
geolocations of the \acp{BSSID} the client submitted. 

Apple's \wifi geolocation API~\cite{applegeo} works in the latter manner, but with
an added twist: 
In addition to the geolocations of the \acp{BSSID} the client
submits, Apple's API opportunistically returns the geolocations of up to
\emph{several hundred more} \replaced{BSSIDs nearby the one requested.}{nearby \acp{BSSID}.}
These unrequested \ac{BSSID} geolocations are presumably then cached by
the client, which no longer needs to request the locations of the
nearby \acp{BSSID} it may soon encounter\added{, \eg, as the user walks
down a city street.}

Figure~\ref{fig:wps} demonstrates the operation of Apple's \ac{WPS}. A device
that needs to trilaterate its own position based on nearby BSSIDs queries
Apple's \ac{WPS} for the locations of the BSSIDs it detects
(Figure~\ref{fig:wpsquery}); multiple BSSIDs' geolocations can be requested in
the same query. When it knows a BSSID's geolocation, Apple's \ac{WPS}, unlike
Google's, also responds with the locations of up to 400 additional nearby BSSIDs
for the querying device to cache (Figure~\ref{fig:wpsresponse}). 

Apple's \ac{WPS} API \added{is free and places few restrictions on its
use.
It} requires neither an API key, authentication, nor an Apple device;
our measurement software is written in Go and runs on Linux. 
\added{Moreover, Apple appears to make no attempt to filter physically
impossible queries.}
The BSSIDs submitted to the \ac{WPS} need not be physically proximate
to each other nor to the device submitting the query; Apple's \ac{WPS}
will respond with geolocations for BSSIDs on two different continents
in the same request to a querier on a third.

The operation of Apple's \ac{WPS} was first described in
2012~\cite{aguessyrapport} and has since been used in other work to geolocate
\emph{known} BSSIDs---those observed nearby~\cite{isniff} or computed from
IPv6 addresses~\cite{ipvseeyou}. We extend prior work by efficiently discovering
large numbers of previously \emph{unknown} BSSIDs.
\section{Related Work}
\label{sec:related}

There is some prior work that has investigated \acp{WPS} and their
accompanying security and privacy concerns.

The earliest such work is a 2009 study by
Tippenhauer~\etal~\cite{tippenhauer2009attacks} that investigated the Skyhook
\ac{WPS}~\cite{skyhook}. At that time, the Skyhook \ac{WPS}
 was used by Apple devices to calculate their own positions.
They did this by submitting nearby BSSIDs to Skyhook
servers, which would look the BSSIDs up in location lookup tables (LLTs). For
BSSIDs whose locations are known, the Skyhook system would then return the
geolocations of these BSSIDs for the local device to compute its own location.
\replaced{Although}{While} Apple ceased using Skyhook the following year
in favor of using its own \wifi geolocation database~\cite{stopskyhook}, its
devices continue to compute their own locations in essentially the same manner. 

Tippenhauer~\etal demonstrated several attacks against the Skyhook \wifi
geolocation service. First, they used another \wifi geolocation database,
WiGLE~\cite{wigle}, to obtain BSSIDs of \acp{AP} in a geographically-distant
city. They then set up rogue \acp{AP} spoofing the BSSIDs of the remote
\acp{AP}. This had the effect of \replaced{tricking}{convincing} Apple devices in the vicinity of the
spoofed BSSIDs \replaced{into incorrectly believing}{to believe} they were in the remote location. 
They also discovered that they \replaced{could}{can} inject false information into Skyhook's
geolocation database by spoofing the location of Apple devices using rogue
\acp{AP}, while also transmitting from \acp{AP} using BSSIDs unknown to the
Skyhook service. These new BSSIDs were eventually added to Skyhook's LLTs after
broadcasting for some time.

Our attacks leverage Apple's \ac{WPS}, which works in much the same way
as the Skyhook system is described. Unlike Tippenhauer~\etal, we are primarily
concerned with discovering the locations of \acp{AP} worldwide and studying
their movement over time. %

Several years later, Feng~\etal extended Tippenhauer's approach to location
services systems beyond Skyhook, including Google's, Apple's, and Microsoft's
\replaced{\acp{WPS}}{\wifi geolocation systems}~\cite{feng2014vulnerability}. They show that all four
of these services can be tricked into providing the client device with an
incorrect location by setting up rogue \acp{AP} from the location to be spoofed,
while simultaneously jamming nearby legitimate \acp{AP}. 

Boutet and Cunche examined privacy concerns inherent in
\acp{WPS}~\cite{boutet2021privacy}. However, their threat model assumed the
adversary was the operator of the \wifi location provider (\eg, Google, Apple,
or Skyhook), which can learn the location of the device using the system from
the BSSIDs it queries. In contrast, our threat model assumes the adversary is
an unprivileged user of a \wifi geolocation API.

Han~\etal reverse-engineered Google's \ac{WPS}'s
method of operation~\cite{han2022location}. Google's \ac{WPS}
functions differently than Skyhook's and Apple's 
insofar as Google's service attempts to geolocate the device submitting the
query, providing it with only the device's computed position given a list
of BSSIDs from the client. They used this knowledge to improve the effectiveness
of the location spoofing attack introduced in previous work.

\added{One related work used Apple's WPS to geolocate \wifi routers, as does
our own study. In \emph{IPvSeeYou}, Rye and Beverly extracted MAC addresses from a legacy
form of \vsix addresses they obtained in large-scale active Internet
measurements~\cite{ipvseeyou}.  Then, they used data from Apple's WPS, along
with BSSIDs from wardriving databases~\cite{mylnikov,openwifi,openbmap}, as an oracle to
find closely-related BSSIDs (\ie within a few bits from the
\vsix-extracted MAC address). Their observation was that closely-related
BSSIDs are a strong indicator that both MAC addresses belong to different
interfaces on the same device.}

\added{Rye and Beverly's attack differs from ours in several key ways.
First, the authors used Apple's WPS solely to geolocate all-in-one combination
devices that have both a WAN and WLAN interface with MAC addresses at small
offsets from each other. By contrast, our work has no such requirements---its
efficacy relies only on an \ac{AP} being powered on and stationary for long
enough to appear in Apple's WPS. Second, our work focuses on
building a worldwide corpus of \wifi geolocation data for the purpose of
observing location changes over time. By contrast, Rye and Beverly's work is
naturally skewed toward countries with large \vsix deployments, and did not
study the privacy problems that result from BSSID movement over
time---a central focus of our work.\footnote{In fact, their study ignored
movement altogether; whenever they saw a device in multiple locations, they
chose one randomly as the ``canonical'' location~\cite[\S7.3]{ipvseeyou}.}}

\added{In sum,} while prior work is concerned with providing a spoofed location to
unwitting client devices \added{or geolocating \vsix addresses}, our goal in this work is to amass a large-scale,
longitudinal corpus of BSSID locations, and examine the privacy concerns this
raises. The next section describes the range of attacks that can be mounted with
such a corpus.
\section{Mass Surveillance Using \acp{WPS}}
\label{sec:threat}

In this section, we describe a novel way to use a WPS to perform mass
surveillance without any \emph{a priori} knowledge.
We focus the details of our attack on Apple's WPS, but the high-level approach
generalizes to other WPSes, as well.

\subsection{Threat Model} %

We assume \replaced{an unprivileged}{a low-power}, low-resource adversary who has no prior knowledge of
in-use BSSIDs.
The adversary need not have a device from any particular manufacturer (e.g.,
Apple or Google), nor does it need to be geographically proximal to any
specific target.
The central goal of the attacker we consider is to gather location and movement
data about a large number of devices, either globally or pertaining to a
specific region of interest.
Such information gathering can enable a wide range of information-leaking attacks;
we present several such case studies in \S\ref{sec:casestudies}.

\subsection{Mass Surveillance Attack} %
\label{sec:gcc}

We assume that our adversary initially knows nothing about the geographic
locations of any \wifi BSSIDs, and uses the Apple \wifi geolocation API as an
oracle to detect online BSSIDs.

In the most na\"ive variant of this attack, the adversary simply generates
BSSIDs at random and queries the WPS for them.
To get a sense of the likelihood that querying for a random BSSID would result
in returning an actual
device's location, first recall that \added{as MAC addresses,} BSSIDs are 48 bits.
Second, note that WiGLE reports 1.19 billion unique \wifi networks in its
20-year corpus.
Using this as an %
estimate for the number of active APs, this would
result in a probability of 0.0004\% that a random BSSID corresponds to a live
device.

An adversary can drastically reduce the number of queries it has to make by \emph{seeding}
their guesses using the list of allocated \acp{OUI} published by the
IEEE~\cite{oui}.
The IEEE allocates 24-bit OUIs\footnote{OUIs are also referred to by the IEEE as
MAC Address Block Large (MA-L) allocations.} to organizations that need to allocate
MAC addresses to network interfaces, for which they pay a small fee.
MAC
addresses are used in the 802.3 Ethernet, Bluetooth, and 802.11 \wifi protocols
to provide link-layer identifiers. An organization that has been allocated an
OUI can then allocate the remaining 24 bits of the MAC address to individual
network interfaces as they choose. The relationship between organizations and
OUIs is one-to-many; large organizations may have tens or hundreds of OUIs
registered.

Importantly, however, allocated OUIs provide a starting point from which to
test for active BSSIDs, which are MAC addresses used for the wireless interface
on \acp{AP}. At the time of writing, the IEEE has made 34,322 OUI allocations
that are listed in its public OUI database~\cite{oui}.
In addition to the IEEE-assigned OUIs, an attacker can also include the
IEEE-assigned OUIs with \replaced{the U/L bit}{the seventh bit of the
first byte (the so-called Universal/Local (U/L) bit)} set to 1
\added{(see \S\ref{subsec:macaddrs})}. The U/L bit, when 0, indicates
that the MAC address is from an OUI assigned by the IEEE and should be
globally-unique.
All IEEE-assigned OUIs have the U/L bit set to 0. When the U/L bit is set to 1,
however, it indicates that the MAC address is ``locally-administered,'' and the
MAC address should not be considered globally-unique. \acp{AP} will sometimes
use MAC addresses from their assigned OUI with the U/L bit set to 1 when
creating multiple virtual interfaces on the same physical hardware.

\begin{algorithm}[t]
\caption{Mass Information Gathering with \acp{WPS} \label{alg:atk}}

\begin{enumerate}
\item $\mathsf{Geolocated BSSIDS} \leftarrow \emptyset$
\item \textbf{Seed} an initial set of OUIs:

$\mathsf{OUIs} \leftarrow $ IEEE-assigned OUIs (U/L = $\{0,1\}$)

\item Until $\mathsf{OUIs}$ is exhausted:

	\begin{enumerate}
	\item Choose an $o$ from $\mathsf{OUIs}$
	\item Generate $2^{14}$ distinct BSSIDs with OUI $o$
	\item Query the WPS for each random BSSID $b$
	\item For each successful response, comprising the location for $b$ and the locations of up to 400 nearby BSSIDs:

		\begin{enumerate}
			\item Add $b$'s location to $\mathsf{GeolocatedBSSIDs}$.
			\item Add each nearby BSSID's location to $\mathsf{Geolocated BSSIDs}$.
		\end{enumerate}
	\end{enumerate}

\end{enumerate}

\end{algorithm}

With both IEEE-assigned OUIs and their locally-administered variants, there are
68,644 potential OUIs to use to bootstrap BSSID discovery using Apple's \wifi
geolocation API.
\added{This is a dramatic reduction of the search space. A na\"ive attacker
guessing random BSSIDs from the 48-bit MAC address space is in effect guessing
randomly from each of the $2^{24}$ = 16,777,216 possible OUIs. By restricting
random BSSID guessing to IEEE-assigned OUIs and their locally-administered
variants, an attacker narrows the search space by 99.6\%. After this
search space winnowing, the attacker then chooses random BSSIDs to query.}
In our own implementation (\S\ref{sec:methodology}), we query for $2^{14}$
random BSSIDs from each of the 68,644 IEEE-assigned and locally-administered
OUIs.

These queried BSSIDs represent a na\"{i}ve understanding of the allocated MAC
address space that any reasonably-informed attacker could gather with
little-to-no difficulty.
\deleted{Yet the reduction in the search space is considerable; by limiting the random
BSSID generation to only these 68,644 OUIs, the adversary reduces its search
space by 99.6\% (68,644 is 0.4\% of 2$^{24}$).}
As we will see in \S\ref{sec:results}, this leads to rapid discovery of live
devices. 
We will also show that, while there are indeed APs with OUIs that are
\emph{not} from this small set of prefixes, our technique is able to
learn these, as well, because Apple's WPS provides up to 400 nearby
BSSIDs with each successful request.

We summarize how an attacker can perform mass information gathering in
Algorithm~\ref{alg:atk}.

After obtaining a large collection of random BSSIDs' locations, an adversary
can then use that collection to explore specific geographic regions of
interest.
To do this, the attacker can simply identify which BSSIDs are in the
target geographic region (or keep querying with random BSSIDs until it
finds some), and then repeatedly issue queries for the locations of
the nearby BSSIDs it returns, thereby getting another list of nearby
BSSIDs, and so on.

Note that the attack outlined here, and the specific case studies in
\S\ref{sec:casestudies}, require only minimal technical sophistication and can
be carried out by individuals with consumer-grade hardware\added{, with no
additional charges or subscriptions}.
Before evaluating the attack's capabilities, we first consider ethical issues
raised by \acp{WPS} and the ability to conduct these attacks.

\section{Ethical Considerations}
\label{sec:ethics}

Any work that has the potential to track individuals remotely \deleted{by querying an API}
naturally raises ethical questions.  \added{As a preliminary step,} we consulted with our institution's IRB,
which determined that this work is not human subjects research. However, we
believe that our conduct of this work and our aims further adhere to the
principles of beneficence and respect for persons.

To demonstrate the feasibility of our attack, we make use of Apple's
crowd-sourced \ac{WPS} via an API. 
This API is used by Apple products to trilaterate their own location
through the geolocations of nearby AP BSSIDs; our queries mimic those that iOS
and macOS devices routinely make. Apple provides this API free of
charge without the requirement to register for the service or obtain an API key;
furthermore, the use of this API appears to be unrestricted in the query rate or
number of queries allowed. \added{During this study, we issued approximately 30
queries per second; at this rate, we did not encounter rate-limiting or observe
service interruptions or outages. Each API call can itself contain multiple
BSSIDs to query, and in practice, we included 100 BSSIDs per API request.} 

To the best of our knowledge, there is no official
Apple policy restricting the use of this API. While no official documentation
for this API exists, it has been publicly documented in
academic~\cite{ipvseeyou,aguessyrapport} and security community~\cite{isniff}
work over the last decade. Google's \ac{WPS}, which we do not examine, has also
been the subject of privacy research~\cite{han2022location}.

This work identifies the potential for harm to befall owners of \wifi \acp{AP},
particularly those among vulnerable and sensitive populations, that can be
tracked using \acp{WPS}. The threat applies even to users that do not own
devices for which the \acp{WPS} are designed---individuals who own no Apple
products, for instance, can have their \ac{AP} in Apple's \ac{WPS} merely by having
\added{Apple} devices come within \wifi transmission range.

In \S\ref{sec:results} and \ref{sec:casestudies}, we demonstrate the ability to
track both individuals and groups of people over time and space. Because the
precision of Apple's \ac{WPS} is on the order of meters, this allows us to, in
many cases, identify individual homes or businesses where \acp{AP} are located.
Out of respect for user privacy, we do not include examples that could
publicly identify individuals in the case studies we examine in this
work.\deleted{and we
will not publicly release our data.} However, determining the identities of individuals or
groups they are a part of---down to individual names, military units and
bases, or RV parking spots---is eminently possible using the techniques we
describe in this work, and is a major motivating factor for our disclosure of
this vulnerability. \replaced{As such, we have disclosed this vulnerability to
Apple and router manufacturers that feature prominently in
our case studies (\S\ref{sec:casestudies}). We discuss our communication with these companies and
remediation measures they have taken in \S\ref{sec:disclosure}.}{As such, we are currently in the
process of disclosing this
privacy vulnerability to Apple.}

\added{The data collected during this study is stored on the authors' machines
and will not be publicly released. After follow-on work to improve user privacy
has concluded, we will destroy the data.}
\section{Data Collection Methodology}
\label{sec:methodology}

This section describes the datasets we collected; we analyze them in
\S\ref{sec:results} and \S\ref{sec:casestudies}. \added{All of our data was
collected from a single vantage point located in an access network in the
United States.}\footnote{To verify that using only a single vantage point did not
introduce bias, we issued a set of queries for 100 test BSSIDs with known
geolocations from five geographically disparate locations: Mumbai, India;
Melbourne, Australia; Warsaw, Poland; Paris, France; and Washington, DC. 
Apple's WPS returned identical geolocations (to eight decimal places) to each
vantage point.  Therefore, we determined that with the possible exception of
mainland China---which rarely appears in our data---vantage point location
appears not to impact the results returned by the WPS.}

\subsection{Month-Long Longitudinal Dataset}
\label{sec:monthmethod}

While the global corpus collection experiment outlined in \S\ref{sec:gcc}
provides a snapshot of the world-wide distribution of BSSIDs, a fundamental
privacy issue with the existence of \wifi geolocation APIs is the potential for
user movement to be captured as these APIs are queried over extended periods of
time.

To measure the potential for user tracking over time, we sample 10 million 
BSSIDs from the global corpus we amass by randomly guessing BSSIDs within known
OUIs. Then, we query the Apple WPS for these BSSIDs on a daily
basis to detect movements in the BSSID geolocations as reported by the API.

With this knowledge, we then characterize the type and amount of movement in our
sample. We examine the locations from which and to which \acp{AP} (and
potentially their
owners) are migrating. 

The results of our month-long querying the Apple geolocation API for the 10
million sampled BSSIDs are reported in \S\ref{sec:monthresults}

\subsection{Year-Long Longitudinal Dataset}
\label{sec:longitudinalmethod}

Next, we use a longitudinal dataset collected between November 2022 and
November 2023 as the basis for demonstrating several practical privacy attack
case studies in \S\ref{sec:casestudies}. This corpus, which consists of over 2
billion distinct BSSIDs, was additionally focused on various hotspots throughout the year
(\eg, Russia's invasion of Ukraine and the August 2023 wildfires on the
Hawaiian  island of Maui). Its purpose is to highlight the types of practical
privacy attacks a motivated attacker can mount, rather than to demonstrate the
ability to develop a worldwide BSSID geolocation database.

However, collecting BSSIDs from a specific geographic area requires knowledge of
``seed'' BSSIDs located in that area, so that a more comprehensive picture of
the \acp{AP} in that location. We used an initial dataset developed using
methodology similar to that outlined in \S\ref{sec:gcc} to build an initial,
worldwide BSSID view. Then, when we wanted to focus on specific regions, we
first filtered for BSSIDs within that geographic region from our global dataset
before repeatedly querying for our areas of interest.

\subsection{WiGLE Data Validation}
\label{sec:wiglevalidation}

Last, to validate our geolocation collection methodology, we compare the results
of querying Apple's \ac{WPS} with data retrieved from WiGLE.

WiGLE~\cite{wigle} is a 20-year-old crowdsourced \wifi, cellular tower, and 
Bluetooth wardriving project. Wardriving is the term for wireless surveying,
often done while mobile, and typically involves annotating the location the 
wardriver is at when her equipment receives the wireless
signals~\cite{wirelessinfidelity,hurley2004wardriving}. WiGLE contains
records of more than 1 billion \wifi BSSIDs and their geolocations, and is
queryable through a rate-limited API. WiGLE is well-known in the wireless
security community for being a source of \wifi geolocation data, so we compare
the results of querying the Apple \ac{WPS} with WiGLE data for 60,000 BSSIDs
selected from popular \ac{CPE} equipment and IoT vendor OUIs---TP-Link, Roku,
Technicolor, and Vantiva. The 60,000 BSSID sample we selected from WiGLE was 
filtered for BSSIDs that had been updated within the last two months (on or
after 1 October 2023) to ensure relative temporal consistency

Of the 60,000 BSSIDs we retrieved from WiGLE, we removed 178 that were
geolocated by WiGLE to 0,0 latitude and longitude, the so-called ``Null Island''
in the Atlantic Ocean. These were likely due to misconfigured equipment, GPS 
devices without an accurate fix, or some other data corruption. We then
requested the geolocations for the remaining 59,822 BSSIDS from the Apple
\ac{WPS}. Of these, 5,951 (10\%) BSSIDs were unknown to the Apple \ac{WPS},
which responds with a \texttt{-180, -180} latitude and longitude for these BSSIDs. The 
remaining 53,871 (90\%) were geolocatable by the Apple \ac{WPS}. Of these, the
vast majority (52,946 or 98\% of the Apple-geolocatable BSSIDs) were located
within 1 kilometer of the WiGLE geolocation. We discuss BSSID movement in more
detail in \S\ref{sec:results} and \ref{sec:casestudies}. 
\section{Systematic Analysis of a Global Corpus}
\label{sec:results}

In this section we describe the results of our global corpus collection, and
detail a month-long study in which we queried 10 million BSSIDs daily.  

\subsection{Global Corpus Collection}
\label{sec:gccresults}

The attack we demonstrate in this work is the ability to curate a worldwide view
of geolocated BSSIDs, from which we can gain a number of insights, as we discuss
in \S\ref{sec:casestudies}. We build this corpus by querying the Apple \ac{WPS} for random BSSIDs
within IEEE-assigned \acp{OUI} and their locally-assigned analogs. 

Of the
1,124,663,296 BSSIDs for which we queried the Apple \ac{WPS}, only
2,834,067 (0.25\%) were successful---that is, were BSSIDs known by the Apple
\ac{WPS}.
In the event that the \ac{WPS} does not
have a record of a requested BSSID, it returns a \texttt{-180, -180} latitude and
longitude pair. When the Apple \ac{WPS} does have a geolocation for
a BSSID, it returns the latitude and longitude of the BSSID as inferred from
measurements made previously by Apple devices to populate the \ac{WPS} database. 

While only $\sim$3 million BSSIDs we randomly chose from within the 68,644 OUIs
were known by the Apple \wifi geolocation API, we learned the geolocations of
substantially more BSSIDs due \added{to} how Apple's \ac{WPS} operates. When an API request is made
for a BSSID that Apple's \ac{WPS} knows the geolocation of, it additionally returns
\emph{up to 400 additional, nearby geolocated BSSIDs.}

In theory, this opportunistic BSSID caching is desirable feature. An Apple
device that queries for the geolocations of nearby BSSIDs in order to
trilaterate its own location will not need to do so again in the future if the
API responds with many BSSIDs near its location---it need only consult its
cache of geolocated BSSIDs when the user walks down the block. However, to an
attacker, the additional BSSIDs returned by the API offer a wealth of
information and comprise the vast majority of BSSIDs discovered in our
experiment.

While we correctly guessed only 2,834,067 valid BSSIDs at random from our
set of 68,644 known OUIs, we learned 488,677,543 geolocated BSSIDs---172 times
more BSSIDs---from the additional geolocation data that the Apple API returns
(a small number were both requested and learned in the additional BSSIDs
returned from a different request).

\begin{figure}[t]
\centering
    \includegraphics[width=\linewidth]{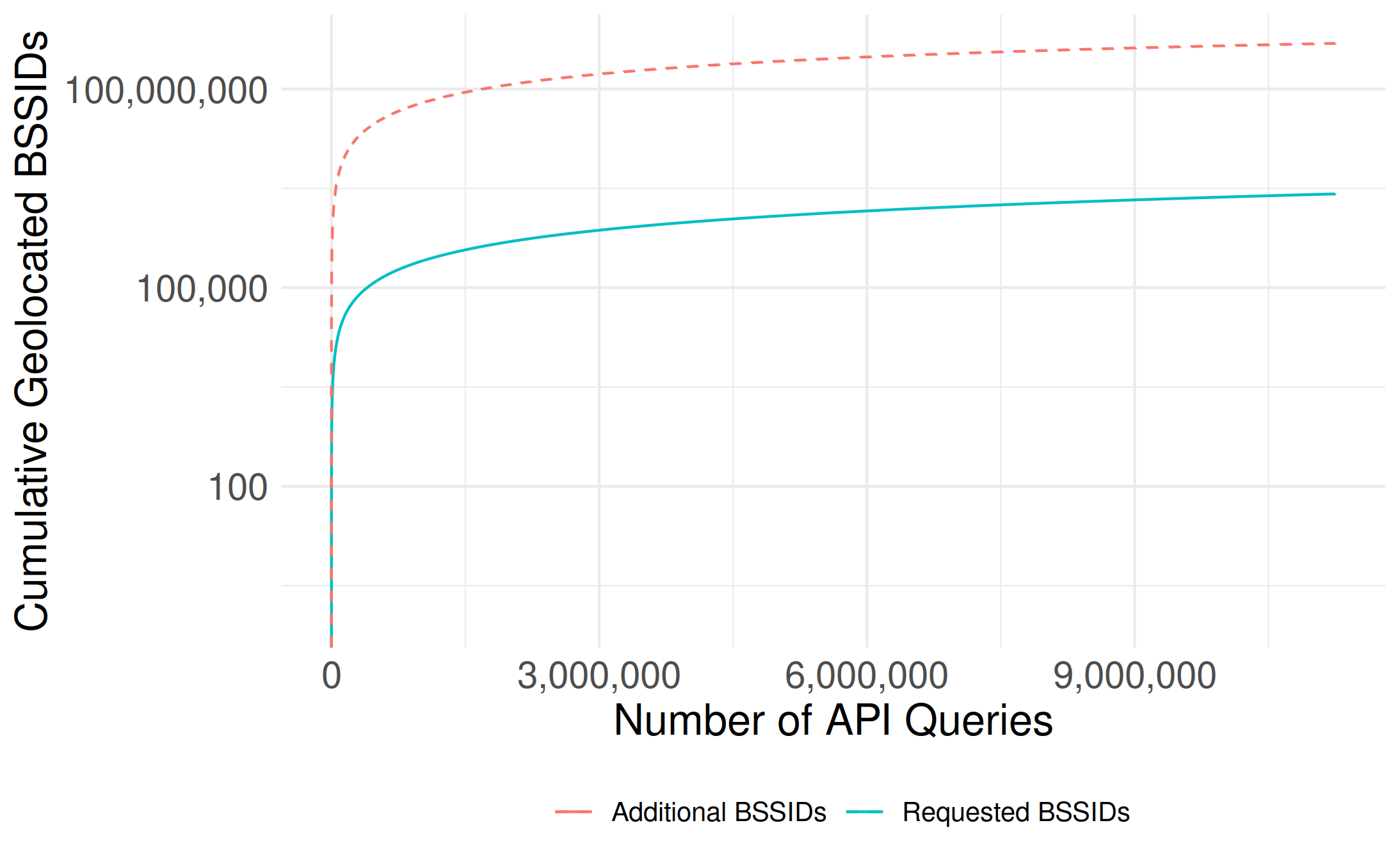}
    \caption{Number of BSSIDs discovered by guessing randomly among
    IEEE-assigned OUIs and their locally-administered versions versus the
    additional BSSIDs the Apple \wifi geolocation API returns. Note that the
    $y$-axis is log-scale.}
\label{fig:guessing}
\end{figure}

Figure~\ref{fig:guessing} displays the cumulative number of BSSID geolocations
 learned from the Apple \ac{WPS} during our global corpus collection. %
The distance between the curves indicate that the knowledge gained from the
additional nearby BSSIDs remains relatively constant as the number of queries
grows \added{(each API query contained 100 BSSIDs to be geolocated)}. Further, the number of new BSSIDs learned continues to grow linearly over
time (Figure~\ref{fig:guessing}'s $y$-axis is log-scale).

\begin{figure}[t]
\centering
    \includegraphics[width=\linewidth]{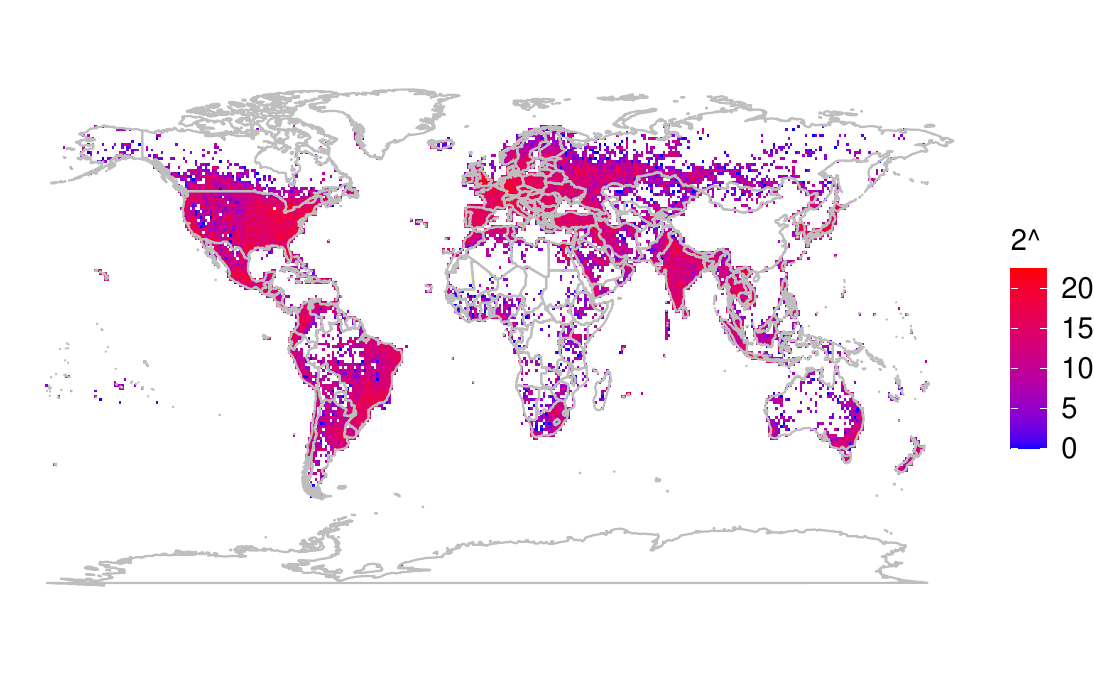}
    \caption{Heatmap of BSSIDs discovered by guessing randomly among
    IEEE-assigned OUIs and their locally-assigned variants.} 
\label{fig:guessing-heatmap}
\end{figure}

The geographic distribution of the corpus we compile through OUI-based random
BSSID guessing is global in scope. Figure~\ref{fig:guessing-heatmap} displays a
global heatmap of the number of unique BSSIDs discovered, aggregated into
4-character geohash bins.  
The distribution of wireless \acp{AP} we discover
largely mirrors the global distribution of human population, with one major
exception. 

China is significantly underrepresented in the BSSID geolocation
corpus, with the exception of the special autonomous regions (SARs) of
Hong Kong and Macau.
We speculate that this is an artifact of Chinese legislation
that restricts collecting and publishing China's (but not their SARs')
geographic information~\cite{china-mapping-law}.
\added{It is possible that, as with some other Apple
data~\cite{china-infrax}, Chinese BSSID location data would be
accessible from a vantage point within China (recall that ours was
in the US).}
In either case, despite the surprising sparseness, we do observe \emph{some}
\wifi \ac{AP} geolocations in mainland China (on the order of
thousands).
We conjecture that these geolocations, which typically arise only for
brief periods of time, are geolocated by tourists or foreigners whose
devices report the geolocation of Chinese \acp{AP}.
That even a nationwide policy not to be included in this dataset can be
so regularly violated underscores the challenges of opting out of a
WPS.

Next, we analyze the BSSIDs we discover by manufacturer, using the first three
bytes (the OUI) of the BSSID to determine the manufacturer. When the
Universal/Locally-Administered (U/L) bit in the BSSID is set, we clear it 
before looking up the OUI, as all IEEE-registered OUIs have the U/L bit set to 0.

\begin{table*}[ht]
    \centering
    \caption{Number of unique geolocated BSSIDs organized by the most
	common OUIs (left) and most common vendors (right).
    Some vendors are listed under multiple names in the IEEE OUI database file;
    we made a best effort to consolidate counts under a single vendor name.}
    \label{tab:random-oui-vendor}
\begin{tabular}{rrcc||rrc}
\multicolumn{1}{c}{\textbf{Count}} &
  \multicolumn{1}{c}{\textbf{\%}} &
  \textbf{OUI} &
  \textbf{Vendor} &
  \multicolumn{1}{c}{\textbf{Count}} &
  \multicolumn{1}{c}{\textbf{\%}} &
  \textbf{Vendor} \\
    \hline
1,119,296   & 0.23  & 12:59:32         & Roku               & 63,628,162  & 12.96 & Unlisted       \\
1,082,808   & 0.22  & 8e:49:62         & Roku               & 42,351,581  & 8.63  & TP-Link        \\
998,529     & 0.20  & 1c:3b:f3         & TP-Link            & 28,407,586  & 5.79  & Huawei         \\
968,367     & 0.20  & 40:ca:63         & Seongji Industries & 19,556,964  & 3.98  & Vantiva        \\
965,516     & 0.20  & 00:31:92         & TP-Link            & 17,251,657  & 3.51  & Sagemcom       \\
485,806,036 & 98.95 & 2,401,736 others & --- & 319,744,602 & 65.13 & 15,302 others  \\
\hline
    490,940,552 & 100   & --- & \textbf{Total}     & 490,940,552 & 100   & \textbf{Total}
\end{tabular}
\end{table*}

Table~\ref{tab:random-oui-vendor} details the most common OUIs and equipment
manufacturers in our corpus obtained from randomly guessing among IEEE-allocated
OUIs and their locally-assigned variants.

\begin{figure}[t]
\centering
    \includegraphics[width=\linewidth]{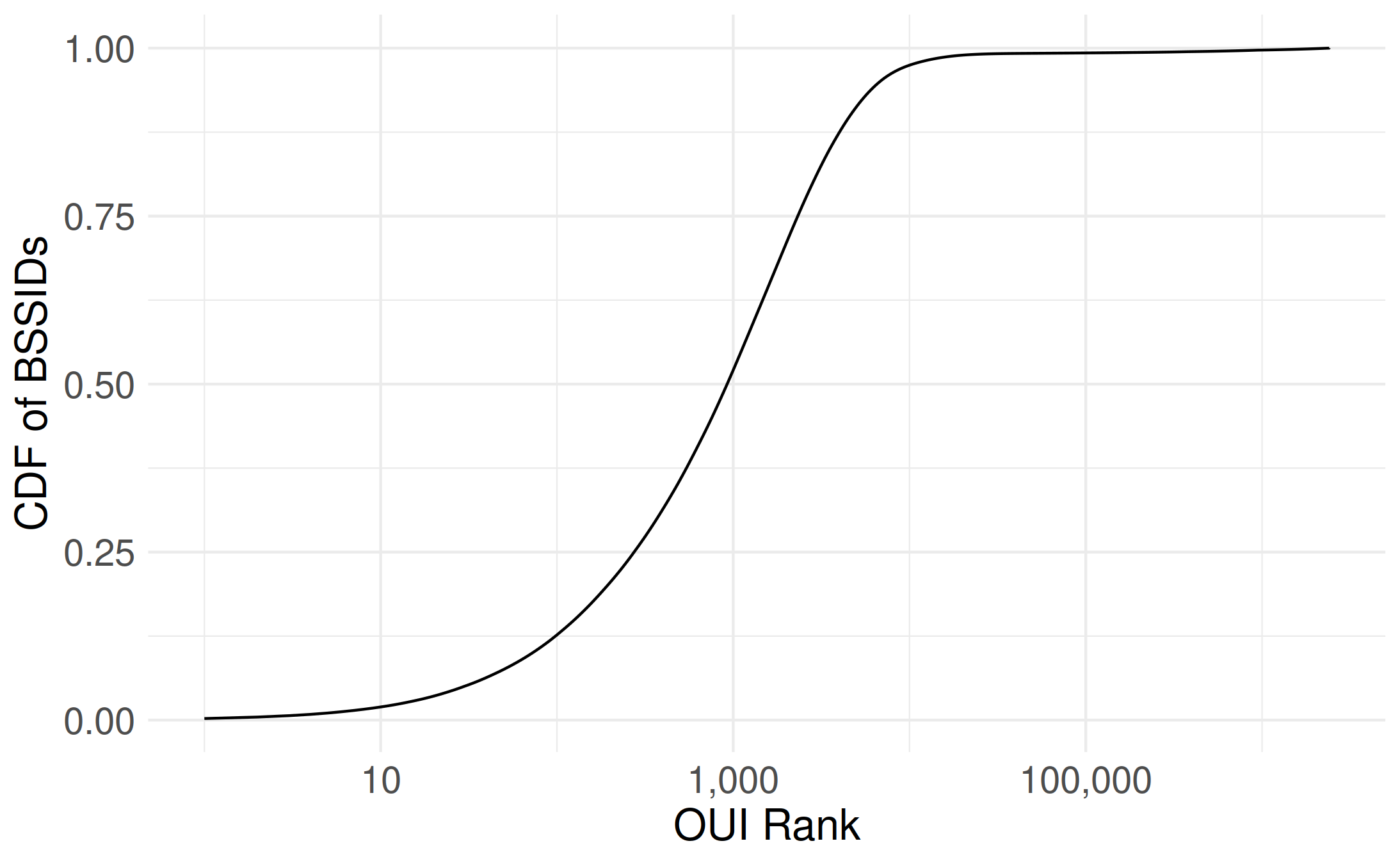}
    \caption{CDF of geolocated BSSIDs versus OUI ranked by decreasing number of
    geolocated BSSIDs. Note that the $x$-axis is log-scale.}
\label{fig:bssid-per-oui-cdf}
\end{figure}

Two of the five most commonly-observed OUIs from the BSSIDs we obtained belong
to the streaming television equipment manufacturer Roku, while another two are
assigned to the home-router manufacturer TP-Link. Over 2 million unique ``OUIs''
were observed in our data; because only about 30,000 OUIs have been assigned by
the IEEE, the vast majority of OUIs in our are likely mobile phones. Mobile
devices running iOS 17 or Android 14, current at the time of writing, choose a
random BSSID when put into hotspot mode. This is further supported by the fact
that most of the ``OUIs'' that do not resolve to a manufacturer have only one
BSSID observation. Figure~\ref{fig:bssid-per-oui-cdf} shows that more than 90\%
of all BSSIDs in our corpus belong to one of the top 10,000 OUIs, while the
remaining $<$10\% is accounted for in the more than 2 million other OUIs.

Next, we aggregate the number of BSSIDs by OUI manufacturer. In some cases, this
is a manual and error-prone process, as manufacturers may specify whatever name
they wish when they register new allocations with the IEEE, and variations in
company names exist between different allocation registrations. Regardless, we
made an attempt to normalize and aggregate discovered BSSID by manufacturer on
the right side of Table~\ref{tab:random-oui-vendor}. The most common vendor our
BSSIDs mapped to was no vendor at all, meaning that the BSSID's OUI had no
listed manufacturer. Again, we believe these are primarily mobile phones, which
choose a random BSSID when placed into hotspot mode in order to prohibit device
tracking. While the most common vendor, this occurred for only about 13\% of BSSIDs.

The most common true vendors in our dataset were TP-Link, Huawei, Vantiva, and
Sagemcom. All four of these companies manufacturer \ac{CPE}-grade routers for
home and small business purposes. No single manufacturer accounts for more than
10\% of the dataset, with TP-Link accounting for the largest fraction of the
geolocated BSSIDs at 8.6\%.

\subsection{Month-Long Longitudinal Study}
\label{sec:monthresults}

\replaced{The privacy threats of massive \wifi geolocation datasets are
arguably most severe when wireless \acp{AP} move.}
{Fundamental to this work is detecting and tracking movement among wireless
\acp{AP} over time.}
To determine the prevalence of \ac{AP} movement and
characterize the phenomenon, we randomly sampled 10 million geolocated BSSIDs
\added{from} the corpus of 490 million \deleted{geolocated BSSIDs} we amassed in
\S\ref{sec:gccresults}. 

Over the course of a month, we queried the Apple \ac{WPS} for these
10 million BSSIDs each day in the same order at approximately the same time.
This month-long study allows us to answer questions about the longevity of BSSIDs
in our corpus, in addition to learning how \acp{AP} move over time and where
they move from and to.

\begin{figure}[t]
\centering
    \includegraphics[width=\linewidth]{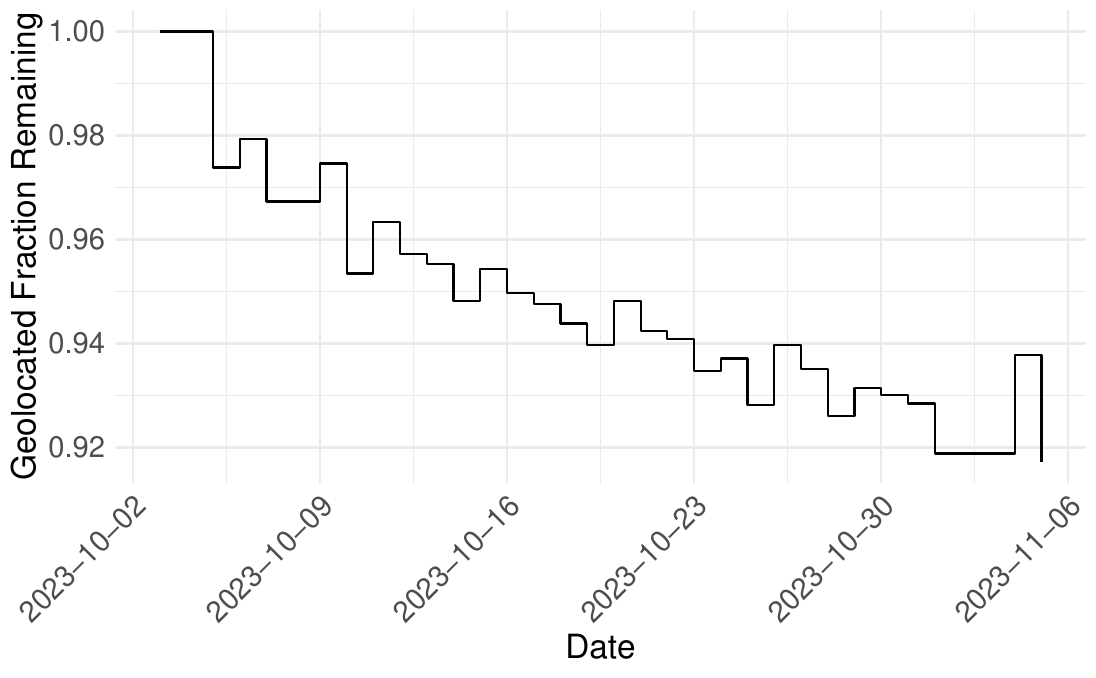}
    \caption{Fraction of geolocated BSSIDs (of 10,000,000 tested) remaining each
    day following an initial sweep of the OUI space.}
\label{fig:day-hits}
\end{figure}

Figure~\ref{fig:day-hits} displays the fraction of the original 10 million that
are still geolocatable each day that we query the API. This figure demonstrates
several interesting artifacts of Apple's \ac{WPS}. First, between the
beginning and end of the month, approximately 8\% of the sampled BSSIDs were no
longer able to be geolocated (that is, the API returned a \texttt{-180,-180}
latitude and longitude for the BSSID). While we do not possess ground truth for
any of the devices in our 10 million BSSID sample, we suspect that there are
several factors that contribute to the decay in geolocatable BSSIDs.

First, some devices are turned off, whether temporarily or permanently, for a
period of time long enough to be reflected in Apple's \ac{WPS}.
We demonstrated in a laboratory setting that it takes roughly a week
for a device to appear in the geolocation database when first powered on, as
well as to be expunged from the database when powered off. It is possible that
some devices have simply been off for long enough to be removed from the
location API because they have been replaced, destroyed, or powered off due to a
long-term power outage.

It is possible that some devices, while still powered and active, are no longer
near enough Apple devices to have their location reported and are therefore
``aged out'' at some point. While we believe that removing BSSIDs from the
geolocation service for this reason is possible, due to the remoteness of some
BSSIDs we discover (\eg, in Antarctica or Tristan Da Cunha Island), we believe
that the threshold for number of ``finding'' Apple devices for a BSSID to be
entered into the geolocation service to be relatively low.

Finally, we suspect that some BSSIDs in the geolocation database are random MAC
addresses, which are used by both modern Android and iOS devices when users put
them into hotspot mode. These BSSIDs are identifiable because they have the U/L
bit set to indicate that they are locally-administered, and when the bit is
unset, the resulting OUI does not map to any assigned by the IEEE. 

We also note that some BSSIDs ``drop out'' of the Apple
\replaced{\ac{WPS}}{\wifi positioning
system}, only to later ``return'' and be geolocatable again. These events
manifest as rises in the fraction geolocatable in Figure~\ref{fig:day-hits}. We believe that
these are \wifi \acp{AP} that, for whichever of the preceding reasons, have been
removed from the \replaced{\ac{WPS}}{\wifi positioning system}. Assuming that the device has not been
physically destroyed, it stands to reason that it can again be plugged in and
used as a geolocation landmark---perhaps after being sold to a new owner, or
after power has been restored following a power outage that kept it offline
for an extended period of time. Further, certain classes of \ac{AP}---travel
routers---are designed to move with the user, which implies periods in which
they are off and not present in Apple's \ac{WPS}. 

\begin{figure}[t]
\centering
    \includegraphics[width=\linewidth]{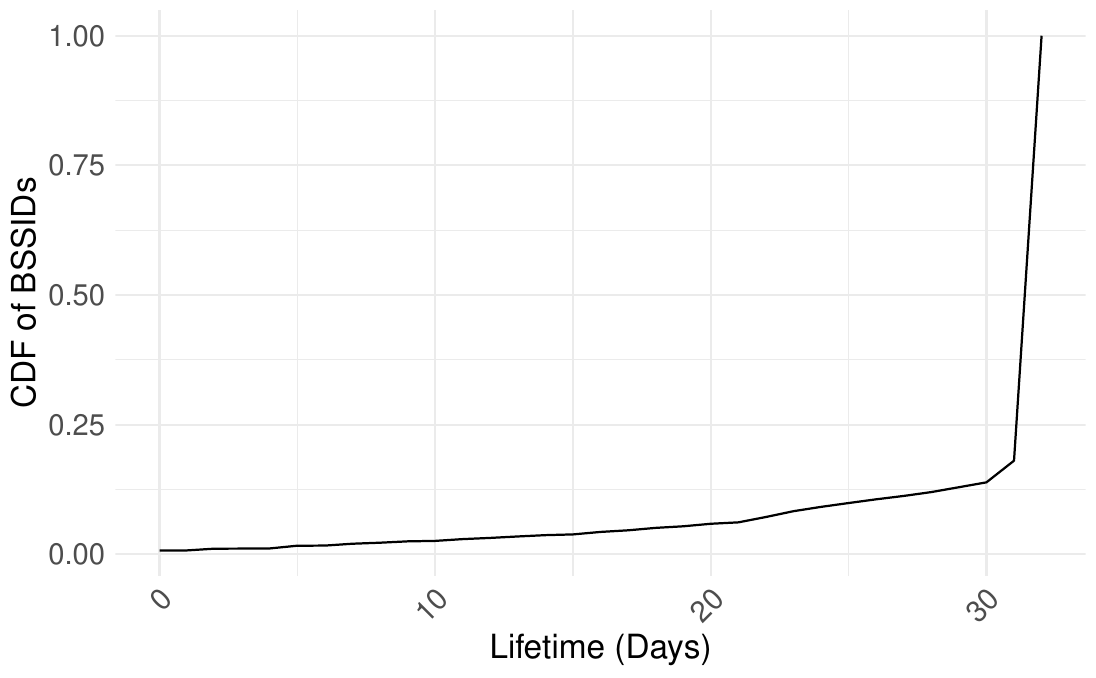}
    \caption{Distribution of lifetimes of geolocated BSSIDs from daily querying
    for 10 million randomly sampled BSSIDs.}
\label{fig:lifetime-cdf}
\end{figure}

Figure~\ref{fig:lifetime-cdf} displays the lifetimes of the 10 million BSSIDs
that were part of our month-long longitudinal study. The lifetime of a BSSID is
the total number of days it was geolocatable, even if one or more breaks
occurred over the course of the month. More than 85\% of BSSIDs remain
geolocatable for the entire month, indicating that most of the BSSIDs in the 10
million BSSID sample are stable.

Next, we characterize the amount of movement we see in our 10 million BSSID
sample. Changes in the locations of the BSSIDs reported by the Apple \wifi
geolocation service is indicative of the \ac{AP} the BSSIDs are tied to changing
geographic locations, and are potential indications that the owner of the
\ac{AP} has moved. This movement is at the heart of the privacy vulnerability
that \wifi geolocation services present, and underlies all of the threats we
present in \S\ref{sec:threat} and demonstrate in \S\ref{sec:casestudies}.

Most BSSIDs in our geolocation corpus change positions over time. These changes
are typically small and are probably due to updates in Apple's computation of
the \ac{AP}'s position.  Therefore, we apply a conservative filter of 1
kilometer total distance moved during the month-long longitudinal experiment to
identify BSSIDs that have moved.

\begin{figure}[t]
\centering
    \includegraphics[width=\linewidth]{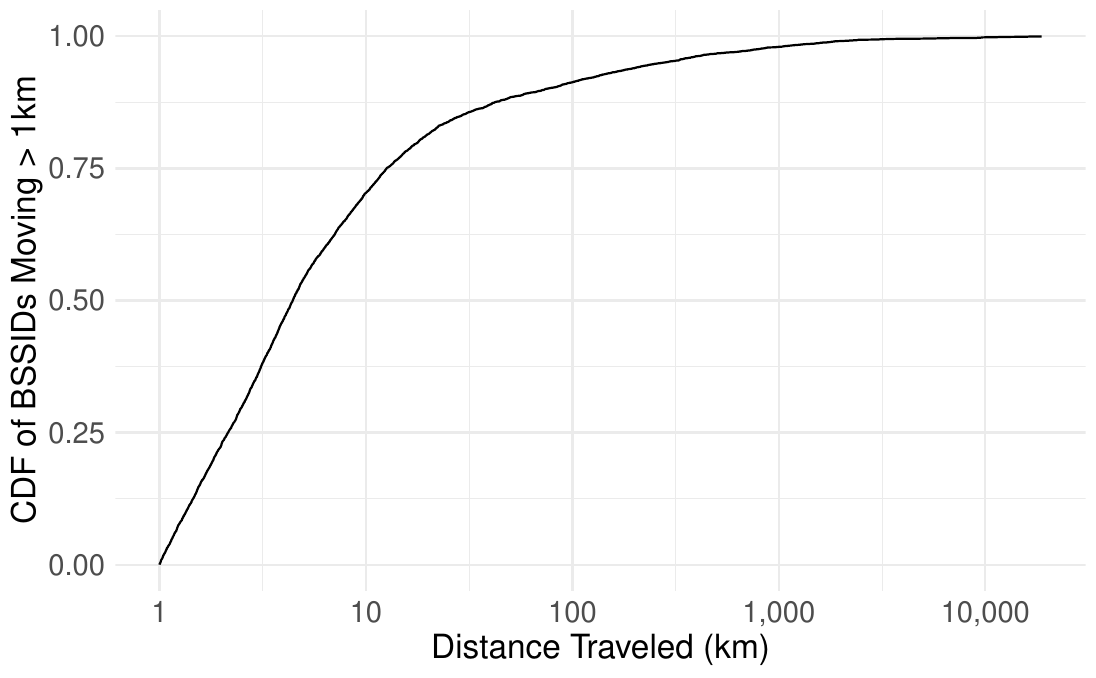}
    \caption{Distances traveled by BSSIDs from the 10 million BSSID
    sample; BSSIDs that traveled less than 1 kilometer are excluded ($N=6,002$).}
\label{fig:dist-cdf}
\end{figure}

Of the 10 million total BSSIDs in our sample, 6,002 move more than 1 kilometer
during our month-long experiment. While 0.06\% is a small percentage of our
initial sample, this indicates that many thousands of BSSIDs move when
considering a larger dataset (such as our corpus from randomly-guessing BSSIDs
within OUIs) over a longer time frame than a month. Figure~\ref{fig:dist-cdf} is
a CDF of the distance BSSIDs in the 10 million BSSID sample travel, restricted
to only those that travel more than 1 kilometer. The median distance traveled is
approximately 4.5 kilometers.

In \S\ref{sec:casestudies}, we delve deeper into some case studies that
illuminate the privacy issues inherent with \replaced{\acp{WPS}}{\wifi
positioning systems}, all of which involve BSSIDs that move over time,
or disappear from the Apple \replaced{\ac{WPS}}{\wifi positioning
system} entirely.
\section{Case Studies}
\label{sec:casestudies}

In this section, we demonstrate how the results from longitudinal \ac{WPS}
queries can be used by an adversary to track \wifi \acp{AP} and their owners over
time. In contrast to \S\ref{sec:results}, in this section we use a corpus of
BSSID geolocations obtained from the Apple \ac{WPS} between November 2022 and
November 2023. This corpus contains more than 2 billion unique geolocated
BSSIDs. Much of the data collection in this corpus was focused on
either specific geographic regions or OUI vendors. To collect geolocation data
specific to a region, we filtered known BSSIDs from a global corpus that were
within a specified geographic bound (\eg, Hawaii or Ukraine). With this initial
``seed'' set of BSSIDs within the geographic area of interest, we then queried
Apple's \ac{WPS} for these BSSIDs, from which we learned more BSSIDs within the
geographic region due to how the Apple API returns up to an additional 400
geolocated BSSIDs nearby the queried BSSIDs. We repeated this process until we
learned most of the BSSIDs the Apple \ac{WPS} knows for a geographic region.

\subsection{Maui Wildfires}
\label{sec:maui}

The Hawaiian island Maui was devastated by a large fire in August 2023, which
destroyed much of the town of Lahaina. During
the wildfires, we queried Apple's \ac{WPS} for BSSIDs we had previously observed
in and nearby Lahaina. Many BSSIDs were ``forgotten'' by the Apple \ac{WPS} 
after a period of several days, presumably after they had been destroyed by
the fire.

\begin{figure}[t]
\centering
    \includegraphics[width=\linewidth]{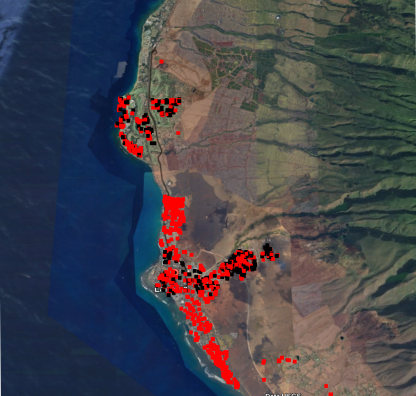}
    \caption{A view of damage from the August 2023 Maui fires. Red points are
    BSSIDs that were in Lahaina, Hawai'i prior to the fire but were not
    geolocatable in the Apple \ac{WPS} as of mid-October. Black points were
    geolocatable prior to and one month after the fires.}
\label{fig:lahaina}
\end{figure}

Figure~\ref{fig:lahaina} depicts BSSID from Lahaina before and after the fire. Black dots are
BSSIDs that were in the Apple \ac{WPS} before the fire started as well a month
after. Red dots indicate BSSIDs that were removed from the Apple
\ac{WPS} between the fire starting and mid-October. We suspect that most of the
\acp{AP} associated with these BSSIDs were destroyed in the fire. This is not
purely speculation; our map of BSSIDs that disappeared from the Apple \ac{WPS}
substantially aligns with open-source reporting and maps of the damage produced
following the fire~\cite{nytmaui}.

This case study demonstrates that real-world
phenomena manifest in the data we retrieve from the Apple \ac{WPS}. In
addition to studying the effects of natural disasters like fires and hurricanes,
a remote attacker might also use an \ac{WPS} in this manner to conduct a battle
damage assessment following a military strike or terrorist attack.

\subsection{User Tracking}
\label{sec:tracking}

In \S\ref{sec:results}, we
found that by querying 10 million BSSIDs daily only 0.06\% moved more than one
kilometer over the course of a month. Many of the \ac{AP} vendors from that
sample manufacture infrastructure that seldom moves---both commercial and
residential \wifi deployments are rarely taken down and set back up once they
are installed. 

However, certain types of \acp{AP} \emph{are designed for mobility}. For
example, the router manufacturer GL.iNet~\cite{glinet} produces a variety of
small ``travel routers'' designed to be used for \eg, in hotels, boats, and
recreational vehicles.

Of 511,935 GL.iNet BSSIDs we geolocated at any point over the yearlong corpus
collection, 23,396 BSSIDs moved more than 1 kilometer, our conservative
threshold to detect \ac{AP} mobility. This 4.6\% BSSID movement rate is
76 times greater than the mobility observed in our 10 million randomly
sampled corpus, and highlights the different use cases of different router
types.

\begin{figure}[t]
\centering
    \includegraphics[width=\linewidth]{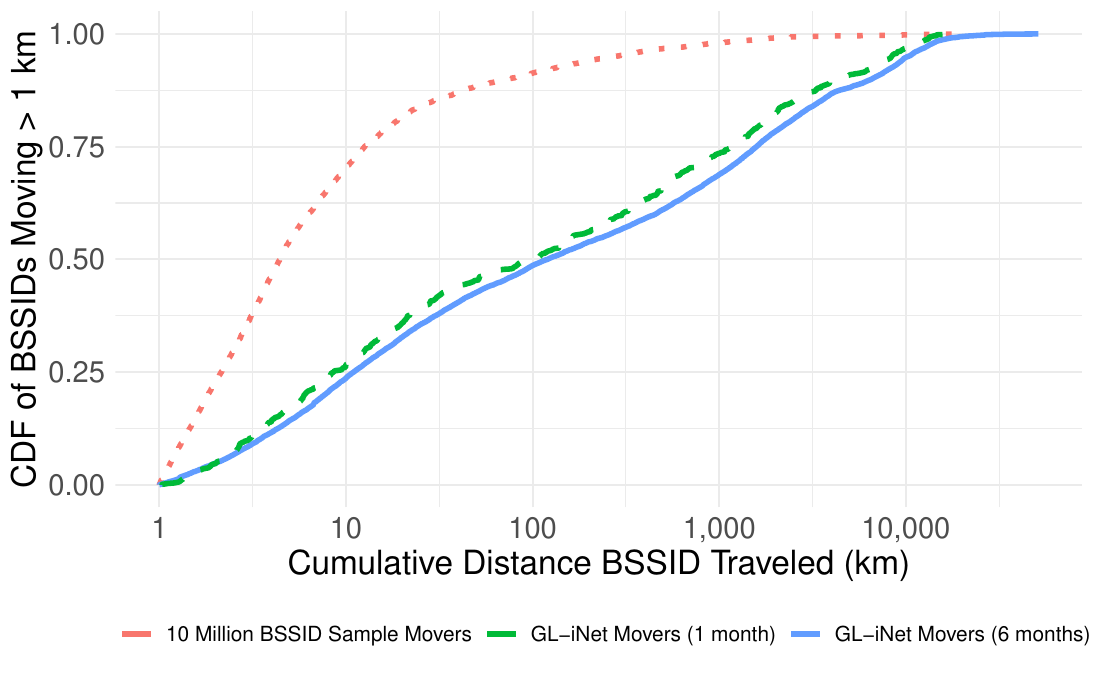}
    \caption{Comparison of the cumulative distances traveled between
    BSSIDs that moved in our 10 million BSSID sample, GL.iNet travel routers
    restricted to the same timeframe as the 10 million BSSID sample, and GL.iNet travel
    routers over six months. $x$-axis is log-scale.}
\label{fig:movercompare}
\end{figure}

Figure~\ref{fig:movercompare} shows that the GL.iNet routers move significantly
farther than the movers in the month-long sample. While $\geq 1$ kilometer moving routers in the 10 million
BSSID sample have a median distance traveled of 4 kilometers, moving GL.iNet routers
over the same time period traveled a median 97 kilometers. Over the six months
during which we tracked GL.iNet travel routers, the median distance traveled
grew to 120 kilometers.

This suggests that when general-use routers move, such as those rented or lent to
customers by \acp{ISP}, their movements are less pronounced than routers that
are specifically designed for movement. While we do not have ground
truth, we surmise that these types of routers, when they do move, are often
redistributed to customers of the same \ac{ISP} that live in the same general
area from a local customer support office.

While we omit specific details to protect the users we were able to track,
movement from BSSIDs represents a serious privacy problem. We observe routers
move between cities and countries, potentially representing their owner's
relocation or a business transaction between an old and new owner. \added{While
there is not necessarily a 1-to-1 relationship between \wifi routers and
users, home routers typically only have several.} If these
users are vulnerable populations, such as those fleeing intimate partner violence or a
stalker, their router simply being online can disclose their new location.

Travel routers, \added{like many GL.iNet devices}, compound the problem. Because travel routers are
frequently used on campers or boats, we see a significant number of GL.iNet
devices move between campgrounds, RV parks, and marinas. They are used by
vacationers who move between residential dwellings and hotels. We have
evidence of their use by military members as they deploy from their homes
\added{and bases} to war
zones.

The ability to detect router movement is a grave threat to individuals who wish
to not be tracked. In the next section, we discuss privacy threats to entire
populations of sensitive populations and vulnerable people -- military members
and civilians living through the wars in Ukraine and Gaza.

\subsection{Russia-Ukraine War}
\label{sec:ukraine}

The ongoing Russian invasion of Ukraine provides further examples of privacy
threats presented by \acp{WPS}. 

First, we examine where BSSIDs that are seen in Russian-occupied Ukranian
territory (the Donbas region in the east and the Crimean Pennisula to the south)
originate, if they were seen outside of those two regions. 

\begin{figure}[t]
\centering
    \includegraphics[width=\linewidth]{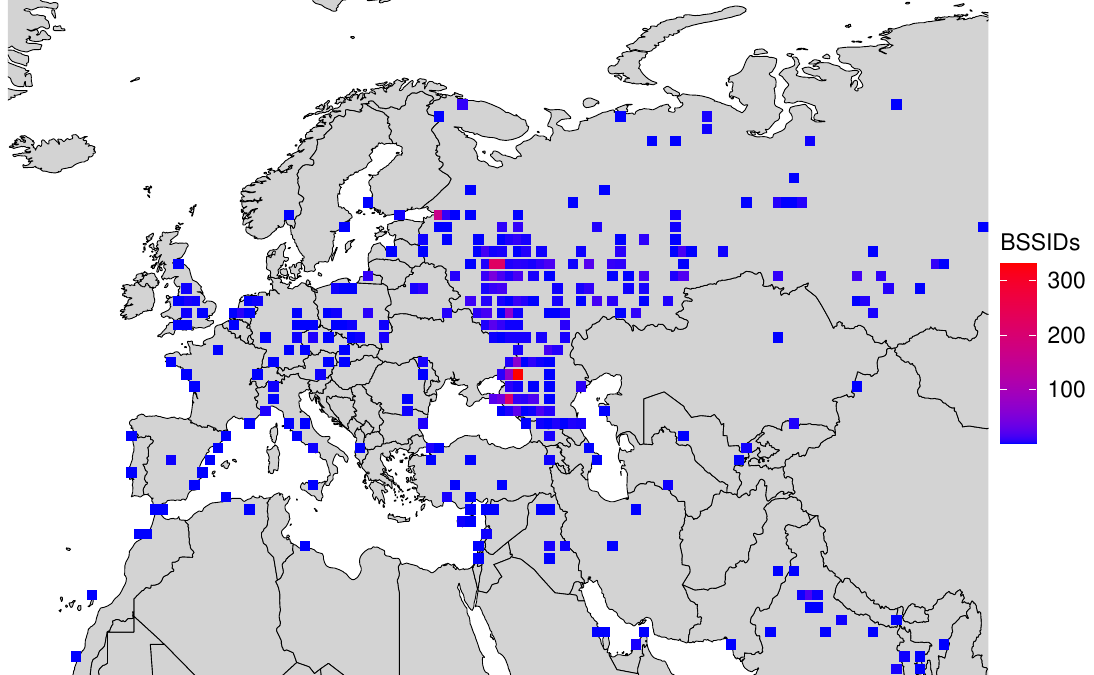}
    \caption{Heatmap of where BSSIDs that enter the Donbas and Crimea regions of
    Ukraine originate.}
\label{fig:uainflows}
\end{figure}

Figure~\ref{fig:uainflows} is a heatmap of the locations that BSSIDs observed in
the Donbas or Crimea were seen \emph{before} they entered Ukraine. Four primary 
Russian hotspots emerge at Moscow, St. Petersburg, Rostov-on-Don, and Krasnodar.
While Crimea remains open for Russian tourism, movement of BSSIDs to these
regions, and especially to the Donbas, is suggestive of military movement. For
BSSIDs not previously in Russia, we speculate that the owners of these devices
are perhaps humanitarian assistance workers, or volunteers with Ukraine's
Foreign Legion. It is also possible that the devices were sold secondhand before
ending up in Ukraine.

Ukraine's use of SpaceX's Starlink satellite Internet service has been widely
reported~\cite{starlinkinvasion,ray2023starlink,itzhak2023russian,henderson2023space}. The \wifi \acp{AP} provided with the Starlink
terminals have BSSIDs drawn from the ``TIBRO Corp'' \texttt{74:24:9F} OUI
(``TIBRO'' is ``ORBIT'' backwards). During the course of our study, we
discovered 1,500,506 Starlink BSSIDs. Of these, 3,722 were geolocated at one
point to Ukraine. 

\begin{figure}[t]
\centering
    \includegraphics[width=\linewidth]{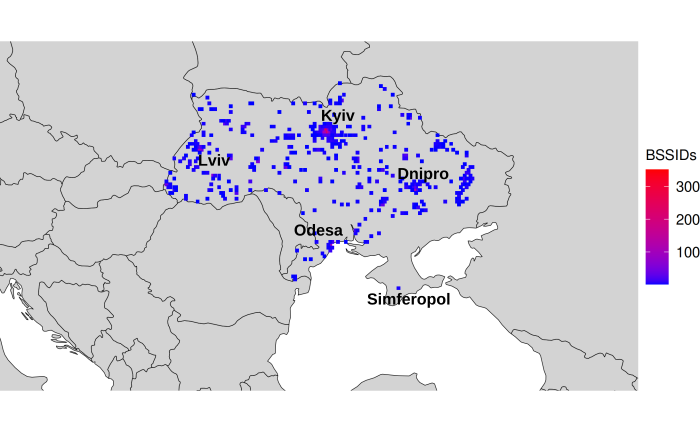}
    \caption{Heatmap of Starlink routers in Ukraine.}
\label{fig:uastarlink}
\end{figure}

Figure~\ref{fig:uastarlink} is a heatmap of the Starlink BSSIDs in Ukraine.
Major hotspots exist in Kyiv, Lviv, Odesa, and Dnipro, the
\replaced{last}{latter} of which is
closest to the front lines in Ukraine's east.

Ukraine's Starlink use is reportedly restricted from functioning on
Russian-occupied Ukrainian territory~\cite{starlinkgeofence}.
Figure~\ref{fig:uastarlink} largely confirms this, with the contours of the
Donbas region's front lines evident east of Dnipro.

Interestingly, one Starlink BSSID did appear in Russian-occupied city of
Simferopol, Crimea during
our study. However, it is worth noting that the Starlink \ac{AP} can be
operational without having a connection to the Internet through the Starlink
satellite terminal, which is a potential explanation for this phenomenon.

\subsection{Gaza}
\label{sec:gaza}

On 7 October 2023, \deleted{terrorists affiliated with} Hamas launched an attack on
Israeli settlements from the Gaza Strip. In response, Israel
\replaced{initiated}{retaliated with} a
major air and ground \replaced{war}{campaign} in Gaza.

We are able to view the effects of Israel's war in Gaza through the lens of
geolocated BSSIDs in Apple's \ac{WPS}. Shortly after the Hamas
attacks, Israel ceased providing electricity to the Gaza Strip. While some fuel
aid is allowed into Gaza by relief organizations, the Israeli power cut severed
a large fraction of the power used by Gazans to power basic household items,
like \wifi \acp{AP}~\cite{gazablackout}.

Beginning shortly after the Hamas attack on Israel, we queried for both Gazan and
Israeli BSSIDs to understand the effects of the \replaced{Israel-Gaza War}{Gazan power outage}. Each day, we
queried about 300,000 Gazan BSSIDs to see whether they are still known to Apple's
\ac{WPS}, and as a control, we queried a similar number of BSSIDs
from a neighborhood in Tel Aviv, Israel.

\begin{figure}[t]
\centering
    \includegraphics[width=\linewidth]{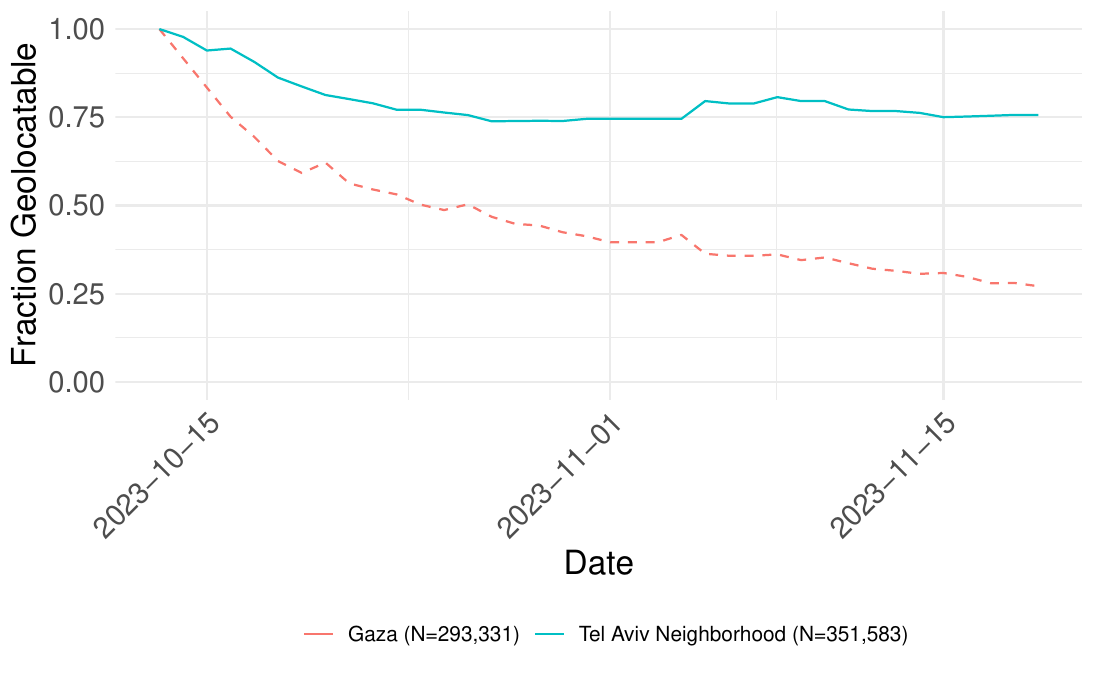}
    \caption{Comparison of the decay in geolocated BSSIDs in Gaza and Tel Aviv,
    Israel.}
\label{fig:israel-decay}
\end{figure}

Figure~\ref{fig:israel-decay} demonstrates the effects of the Gaza power cut.
Within one week, the percentage of Gazan BSSIDs that were geolocatable at the
outset of the Israel-Gaza war had dropped to 62\%. In contrast, the number of
Tel Aviv BSSIDs that continued to be geolocatable was nearly 20\% higher, with
81\% still known to Apple's \wifi geolocation API.

As time progressed, the number of Gazan BSSIDs that are geolocatable continued
to decline. By the end of the month, only 28\% of the original BSSIDs were still
found in the Apple WPS. In contrast, the number of BSSIDs
in the Tel Aviv control group leveled off after initially declining, and stayed
relatively consistent after an initial drop within the first 10 days. At the
end of the month, 76\% of the Tel Aviv BSSIDs were still geolocatable, having
lost only an additional 5\% since the end of the first week.

\begin{figure}[t]
    \begin{subfigure}[c]{0.48\textwidth}
    \includegraphics[width=\linewidth]{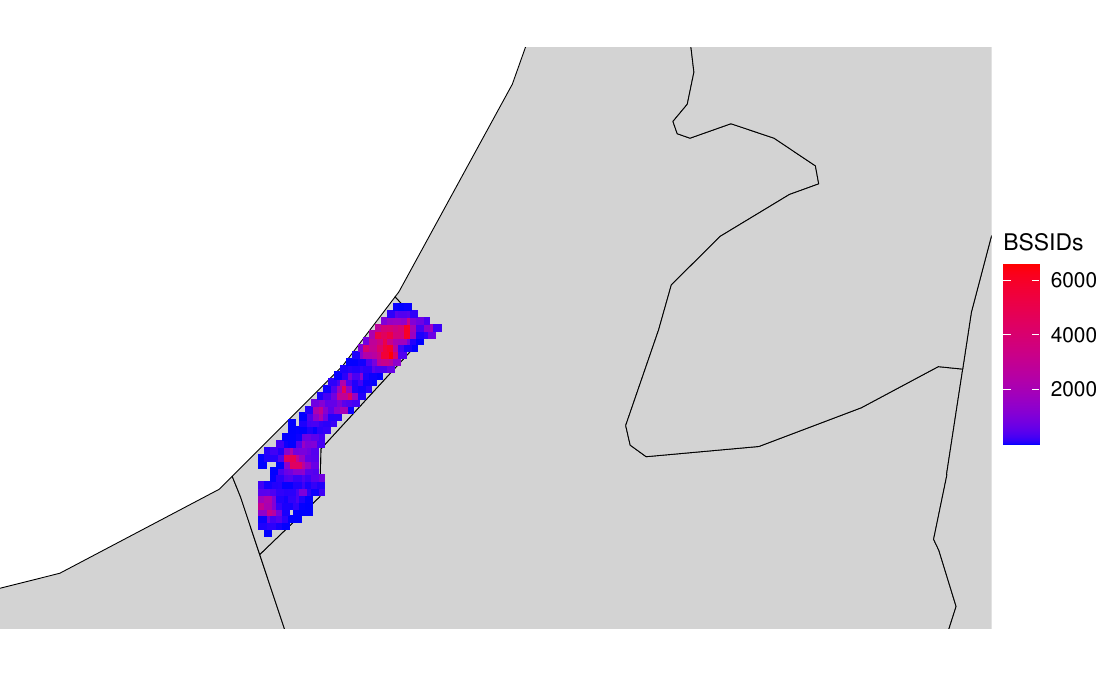}
    \caption{Heat map of the BSSIDs in Gaza on 13 October 2023. Several hotspots
        are present in dense urban environments.}
    \label{fig:gazafirst}
  \end{subfigure}
    \hspace{1em}
    \begin{subfigure}[c]{0.48\textwidth}
    \includegraphics[width=\linewidth]{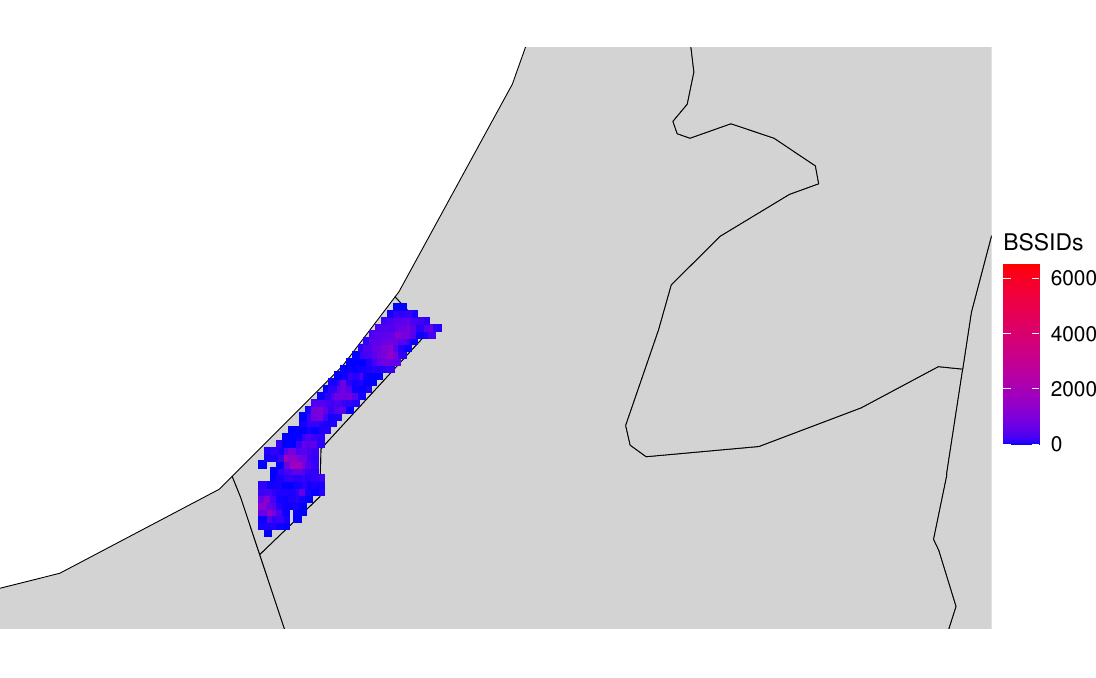}
        \caption{Heat map of the BSSIDs in Gaza on 19 November 2023.
        Significantly lower density is seen in previous hotspots.}
    \label{fig:gazalast}
  \end{subfigure}
    \caption{Heatmaps of BSSID density within the Gaza strip following the
    attack on Israel and Israel's response.}
    \label{fig:gazaheatmaps}
\end{figure}

Figure~\ref{fig:gazaheatmaps} displays the drop-off in geolocatable BSSIDs by
geographic region inside Gaza. On 13 October 2023 (Figure~\ref{fig:gazafirst}), after
the initial Hamas attack and before the beginning of the Israeli ground
war, major hotspots of dense BSSID geolocations exist, but are most
pronounced in northern Gaza. Figure~\ref{fig:gazalast} depicts the BSSID
density on 19 November 2023. Significantly weaker hotspots still exist, but
northern and southern Gaza's BSSID density has largely normalized.

Finally, we considered the BSSIDs that were in Gaza on 7 October---when the war
between Hamas and Israel began---that moved more than 1 kilometer. Only 14
BSSIDs were detected on or after 7 October within Gaza that were detected again
after moving. Of these, the majority (10) moved within Gaza. Several moved from
North Gaza, where the initial Israeli offensive occurred, to South Gaza, where
people living in the north were instructed by the Israelis to move at the war's
outset. One BSSID moved from Gaza to each of Romania, Poland, C\^{o}te d'Ivoire,
and Israel. Note that because BSSIDs linger in the \ac{WPS} for several days
after being taken offline, it is possible that some of the BSSIDs detected on 7
October had already begun moving by that date. Nevertheless, the timeframe and
movement of the BSSIDs from north to south within Gaza suggests that this
relocation was in response to the war.

\section{Remediation}
\label{sec:remediation}

There are policy-based, technical, and legal solutions to remediate the privacy
concerns created by the presence of large-scale \wifi geolocation APIs. They
range in effectiveness from preventing only a weak, non-state adversary from
performing global tracking of \wifi \acp{AP} while permitting targeted tracking,
to eliminating most \ac{AP} tracking if implemented correctly.

\parhead{\wifi \replaced{Positioning System}{Geolocation Service} Remediations}
The type of unlimited and unregulated access to \wifi geolocation data Apple
currently permits via its API should be prohibited. \deleted{While imperfect, Google's
\wifi geolocation API offers an example of commonsense controls that can and
should be applied to geolocation APIs.} \added{There are a number of controls
that Apple might implement in order to restrict access to this data.}

\added{Perhaps the most straightforward solution to limit information disclosure
is to implement a per-device rate limit. The number of queries legitimate Apple
products issue to the WPS for its intended purpose---self-geolocation---is
likely orders of magnitude lower than an attacker would need to amass a
worldwide BSSID geolocation corpus in a tractable time, especially when combined
with other mitigations (\eg, not returning additional, nearby BSSID locations).
This rate limit could be enforced by modifying the API to tie queries to
specific Apple IDs, which would also allow Apple to ban or further limit abusive
users.}

To wit, Google's \ac{WPS} requires the use of an API key by which they manage
\replaced{billing for geolocation queries}{billing for the queries for BSSIDs}
that a user performs. The cost is nominal
when the number of queries is small; at the time of writing, Google's
geolocation API query billing rate begins at \$5.00 per 1,000 requests, with
further discounts applied for bulk lookups in excess of 100,000
requests~\cite{googlebilling}.
Attempting to enumerate the allocated OUI space, however, would cost
millions of dollars---likely prohibitively expensive for all but very powerful
adversaries. \footnote{While Google does impose a geolocation API query rate limit,
it is 100 queries per second per user---more than three times higher than
our query rate in this work.}

As noted previously, Apple's \ac{WPS} opportunistically returns up to 400
unrequested BSSIDs and their geolocations that are \replaced{near requested
BSSIDs}{nearby requested BSSIDs whose locations are known}.
\added{This greatly facilitated the breadth and speed at which our attack
was able to operate.}
\deleted{In practice, these locations are then used by the requesting device to
trilaterate its position using the signal strengths of the geolocated BSSIDs
Apple returns.}
We recommend that Apple's \ac{WPS} cease providing unrequested nearby BSSID
geolocations.
While this would not in theory prevent an attacker from amassing
a global BSSID geolocation corpus, it would significantly increase the difficulty,
particularly for random OUIs.
\added{We suspect that limiting the number of returned BSSID geolocations
returned need not adversely affect Apple's geolocation API; after all, Google's \ac{WPS}}
returns only a single geolocation---that of the device that
\replaced{ostensibly}{has} sensed those nearby BSSIDs.
This distinction \replaced{not only}{both} obscures the locations of the BSSIDs in question
by returning the trilaterated requester's position, but also provides nowhere
near the benefit that the Apple API provides in unrequested nearby BSSIDs.
Discontinuing the inclusion of nearby BSSIDs with active
queried BSSIDs would \deleted{also} not prevent an attacker from performing a targeted
tracking attack by querying the API for a specific BSSID or set of BSSIDs
over time.

Finally, the volunteer-sourced wardriving website WiGLE has instituted policy
restrictions associated with sensitive areas. Following the Russian invasion of
Ukraine in 2022, \wifi spottings in Ukraine have been removed from the WiGLE map
and API query results. Operators of other \acp{WPS} could institute similar
policies for sensitive geographic areas, such as war zones. However,
we recognize that implementing such policies might prove problematic for large corporations in
practice. Selectively implementing privacy protections for some areas, while
choosing not to implement them for others, could lead to accusations of bias or
of ignoring the plight of certain populations.

\parhead{\ac{AP} Manufacturer Remediations}
The most comprehensive solution to the range of privacy attacks \wifi
geolocation APIs present that we outline in this paper is to implement MAC
address randomization for \wifi \acp{AP}.

Randomizing the source MAC address field in 802.11 probe requests became
standard among 802.11 client devices in the mid-2010s, with all major operating
systems now implementing this privacy technique. While some implementation issues were
initially uncovered~\cite{matte2016defeating,vanhoef2016mac,martin2017study}, 
\emph{pre-association} MAC address
randomization is now an effective mechanism to prevent tracking mobile 802.11
clients~\cite{fenske2021three}. In 2020, both Android and iOS also began randomizing
clients' source MAC addresses during authentication and after they join an
802.11 network. \emph{Post-association} MAC address randomization prevents
network operators and rogue \acp{AP} from tracking users based on their
hardware MAC address, which was previously exposed any time the device attempted
to join or joined a network.

We are \added{strongly} advocating for a similar privacy measure for wireless \acp{AP}. Wireless
\acp{AP} should choose new BSSIDs each time they are powered on.  Some
\acp{AP} already implement this privacy measure---modern iOS and some Android
mobile devices choose a random BSSID each time they are placed into hotspot
mode. This feature is presumably to safeguard against the type of tracking
attacks mobile device manufacturers have guarded against for nearly a decade
with probe request randomization, and later, post-association client MAC address
randomization. 

Yet, due to the privacy attacks we outlined in \S\ref{sec:threat} and
demonstrated in \S\ref{sec:results} and \ref{sec:casestudies} we believe these
privacy protections should be afforded to all \acp{AP}.
There is no technical \added{or usability} reason why an \ac{AP} should
maintain the same BSSID for long periods of time, or across reboots.
\added{Users would not have to reconfigure their clients;} \wifi clients
typically will associate with any network that advertises an SSID and \ac{AKM}
mode (open, \ac{PSK} or 802.1X \ac{EAP}) they have saved---without respect to
the BSSID.
Indeed, \acp{ESS}, like many campus or corporate networks, rely on the 802.11
client \emph{not} caring about the BSSID in order to transition between
wireless \acp{AP}.

\parhead{User Remediations}
\label{sec:userremediations}
While the average user's privacy is largely at the mercy of \ac{WPS}
operators and the equipment vendor that manufactures their \ac{AP},
there are some practical steps they can take in order to limit their exposure to
the attacks we outline in this work. 

First, users concerned about being tracked via their BSSIDs over time as they
change locations (\eg, moving homes or traveling) should avoid using the same
\ac{AP} at each location if possible. Many home and business subscribers lease
\ac{CPE} equipment from their \ac{ISP}; if given the option to retain their
equipment or turn their equipment in and receive a new device at the
destination, a privacy-conscious user should opt for the latter. 

Second, \acp{WPS} require BSSIDs to exhibit some degree of stability 
before adding them to their database of landmarks. Anecdotally, we find that an
\ac{AP} running in a suburban area will appear in Apple's \wifi geolocation
system \replaced{after two to seven days}{after approximately one week} of continuous operation. While we do not
know the exact mechanics of any \ac{WPS}, \replaced{establishing}{having} some lower
bound on \ac{AP} trustworthiness (whether in terms of time or number of
observers) is intuitive---there is no use navigating by an unreliable
landmark. Users with privacy concerns may thus wish to limit how long 
they use their \ac{AP} in order to prevent it from appearing in a \ac{WPS}.

Finally, the most technically-savvy, privacy-conscious users with the ability to
modify the operation of their \ac{AP} software, like \texttt{hostapd}, should
randomize their BSSID when they operate their \ac{AP}. In a \texttt{hostapd}
configuration file, the \texttt{bssid} directive will change the BSSID used
(presuming hardware support) by the \ac{AP} for that network. By randomizing the
network's BSSID at each boot or time the router moves, the user in effect
achieves BSSID randomization without operating system-level support.

\parhead{Legal Remediations}
\label{sec:legalremdiations}
As Figure~\ref{fig:guessing-heatmap} shows,
China is almost entirely absent from the Apple \ac{WPS}, with the exception of
its SARs Macau and Hong Kong. We do not believe that the Chinese law preventing
the Apple \ac{WPS} from storing Chinese BSSID geolocations foresaw this specific
privacy vulnerability. \deleted{nor do we believe China is a model for protecting user
privacy.} However, this work demonstrates that \ac{WPS} operators largely comply
with the laws of the nations in which they operate, and Chinese law in practice
protects Chinese BSSIDs from being tracked on a large scale, \added{at least from
outside China}.

If other nations or municipalities implement similar laws or policies, we
believe that \ac{WPS} operators will comply with these mandates as well. It is
possible that \ac{WPS} operators might be willing to work with governments to
remove particularly sensitive geographic locations, like war zones or government
and military facilities, or specific OUIs that correspond with sensitive
equipment. Based on our data, we do not believe that any such policies are
currently being implemented.
\section{Responsible Disclosure}
\label{sec:disclosure}

\added{Having given careful consideration to mitigations of our attack in
\S\ref{sec:remediation}, we responsibly disclosed this vulnerability to several
concerned parties. While Apple's WPS clearly permits the global information
gathering we rely on to track large numbers of users around the world, it was
less clear to us who besides Apple should be informed prior to the publication
of this work. All \wifi routers and their owners are potentially affected,
because BSSIDs become entries in a WPS merely by being nearby ``finding''
devices for that WPS (\eg, a neighbor's iPhone might report the BSSIDs of an
\ac{AP}, even when the owner possesses no Apple products herself).}

\added{Because of this ambiguity and large population of potentially-affected
users, we disclosed our findings to Apple and Google as operators of large-scale
WPSes. We also contacted the product security teams for SpaceX and GL.iNet, two
of the router manufacturers highlighted extensively in our case studies in
\S\ref{sec:casestudies}.}

\added{Apple has indicated that they are on track to make several changes to
their WPS in order to better protect user privacy. At the time of writing, they
have given \ac{AP} operators the ability to opt out of inclusion in Apple's
WPS by appending the string \texttt{\_nomap} to a \wifi network's
SSID~\cite{appleoptout}. This change brings it in line with Google's WPS and
WiGLE, which have also excluded SSIDs with \texttt{\_nomap} (Google) and
\texttt{\_nomap} and \texttt{\_optout} (WiGLE) since at least
2016~\cite{googleoptout,wigleoptout}.}

\added{We contacted SpaceX's product security team to inform them of the ability
to geolocate Starlink routers through Apple's WPS, given the sensitivity of the
locations of these devices in the ongoing Russia-Ukraine War. SpaceX's security
team informed us that, while Starlink routers originally used static BSSIDs based on
an assigned OUI, they had begun in early 2023 to roll out software updates that
randomize the BSSID.
These updates are ``being deployed fleet-wide on a region-by-region basis'';
most likely, the devices in our dataset that we identified as Starlink were to
devices that have not yet received the software updates.}

\added{Based on this response, we are encouraged that a major router
manufacturer is beginning to implement BSSID randomization, particularly because
of the security concerns of Starlink users in Ukraine. We hope that other router
manufacturers will follow their example in the near future, and that BSSID
randomization will become the norm rather than the exception.}

\added{Finally, we contacted GL.iNet's product security team due to our case
studies that demonstrate its users can be easily tracked over time and space.
Unfortunately, while they do plan to randomize their \emph{client} MAC addresses,
the GL.iNet's team indicated that 
they have no plans to randomize the products' BSSIDs.}
\section{Conclusion}
\label{sec:conclusion}

In this work, we demonstrated the large-scale privacy threat presented by
Apple's \ac{WPS}. A remote, unprivileged adversary, possessing only the
knowledge of which parts of the MAC address space have been assigned by the
IEEE, can quickly build a corpus of hundreds of millions of geolocated
\acp{BSSID}, spanning all seven continents and extending to even the most remote
corners of the Earth.

The ability to obtain this worldwide view of \wifi \ac{AP} distribution is a
privacy vulnerability. Because people often move with their \ac{AP}, querying a
\ac{WPS} for the same BSSIDs over time reveals when routers---and by proxy
their owners---move. We demonstrated that this attack could be applied to
individual users, such as travel router owners, as they move from location to
location. We also showed that \acp{WPS} could be used to find sensitive
equipment, like Starlink routers in Ukraine.

There are practical steps to take to limit this vulnerability. \ac{WPS}
operators can limit access to their APIs, governments can legislate that their citizen's
devices not be used as geolocation landmarks, and users wary of tracking can be
sure to not use the same \ac{AP} at multiple locations. 

However, the most robust solution to this problem is to implement the same
privacy protections that were implemented in mobile devices in \wifi \acp{AP}.
BSSID randomization at each boot, or when the device changes locations prevents
user tracking even in a world in which \ac{WPS} operators permit open access to
their APIs.
\section{Acknowledgments}

\added{We are grateful for the constructive and insightful feedback provided by
our anonymous reviewers and shepherd, which measurably improved this work. We
would also like to thank Rob Beverly for early feedback, Bill Pugh for feedback
and helping connect us with people at Apple, and the product security teams at
Apple and SpaceX for their prompt attention to our work.
This work was supported in part by NSF grants CNS-1943240 and CNS-2323193.}

\bibliographystyle{plain}
\bibliography{conferences,refs}

\begin{thebibliography}{10}

\bibitem{china-mapping-law}
{Surveying and Mapping Law of the People's Republic of China}, 2013.
\newblock
  \url{https://web.archive.org/web/20170525200020/http://en.nasg.gov.cn/article/Lawsandregulations/201312/20131200005471.shtml}.

\bibitem{glinet}
{GL-iNet}, 2023.
\newblock \url{https://www.gl-inet.com/products/}.

\bibitem{aguessyrapport}
Fran{\c{c}}ois-Xavier Aguessy and C{\^o}me Demoustier.
\newblock {Rapport du projet de fin d'{\'e}tudes Interception des {\'e}changes
  dans une connexion SSL/TLS Application {\`a} l'analyse des donn{\'e}es de
  g{\'e}olocalisation envoy{\'e}es par un smartphone}.
\newblock
  \url{https://fx.aguessy.fr/resources/pdf-articles/Rapport-PFE-interception-SSL-analyse-localisation-smatphones.pdf},
  2012.

\bibitem{applegeo}
Apple.
\newblock {Location Services and Privacy}, 2023.
\newblock \url{https://support.apple.com/en-us/HT207056}.

\bibitem{appleoptout}
Apple.
\newblock {About privacy and Location Services in iOS, iPadOS, and watchOS},
  2024.
\newblock \url{https://support.apple.com/en-us/102515}.

\bibitem{wirelessinfidelity}
Hal Berghel.
\newblock {Wireless Infidelity I: War Driving}.
\newblock {\em {Communications of the ACM}}, 2004.

\bibitem{wigleoptout}
Bobzilla.
\newblock {On \_nomap and \_optout}, 2016.
\newblock \url{https://wigle.net/phpbb/viewtopic.php?t=2330}.

\bibitem{boutet2021privacy}
Antoine Boutet and Mathieu Cunche.
\newblock {Privacy Protection for Wi-Fi Location Positioning Systems}.
\newblock {\em Journal of information security and applications}, 2021.

\bibitem{starlinkgeofence}
Jon Brodkin.
\newblock {Pentagon buying Starlink dishes for Ukraine after funding dispute
  with SpaceX}.
\newblock {\em Ars Technica}, 2023.
\newblock
  \url{https://arstechnica.com/tech-policy/2023/06/pentagon-buying-starlink-dishes-for-ukraine-after-funding-dispute-with-spacex/}.

\bibitem{gazablackout}
Kevin Collier and Rima Abdelkader.
\newblock {Near-total internet and cellular blackout hits Gaza as Israel ramps
  up strikes}, 2023.
\newblock
  \url{https://www.nbcnews.com/tech/internet/internet-blackout-hits-gaza-israel-ramps-strikes-rcna122531}.

\bibitem{nytmaui}
Molly~Cook Escobar, Lauren Leatherby, Scott Reinhard, A~Elena Shao, and Charlie
  Smart.
\newblock {Mapping the Damage From the Maui Wildfires}.
\newblock {\em The New York Times}, 2023.
\newblock
  \url{https://www.nytimes.com/interactive/2023/08/10/us/maui-wildfire-map-hawaii.html}.

\bibitem{feng2014vulnerability}
Jun Liang~(Roy) Feng and Guang Gong.
\newblock {Vulnerability Analysis and Countermeasures for Wi-Fi-based Location
  Services and Applications}.
\newblock \url{https://cacr.uwaterloo.ca/techreports/2014/cacr2014-25.pdf},
  2014.

\bibitem{fenske2021three}
Ellis Fenske, Dane Brown, Jeremy Martin, Travis Mayberry, Peter Ryan, and
  Erik~C Rye.
\newblock {Three Years Later: A Study of MAC Address Randomization In Mobile
  Devices And When It Succeeds}.
\newblock {\em Privacy Enhancing Technologies Symposium (PETS)}, 2021.

\bibitem{googlegeo}
Google.
\newblock {Geolocation API Overview}, 2023.
\newblock
  \url{https://developers.google.com/maps/documentation/geolocation/overview}.

\bibitem{googleoptout}
Google.
\newblock {Control access point inclusion in Google's Location services}, 2024.
\newblock \url{https://support.google.com/maps/answer/1725632}.

\bibitem{googlebilling}
Google.
\newblock {Geolocation API Usage and Billing}, 2024.
\newblock
  \url{https://developers.google.com/maps/documentation/geolocation/usage-and-billing}.

\bibitem{han2022location}
Xiao Han, Junjie Xiong, Wenbo Shen, Zhuo Lu, and Yao Liu.
\newblock {Location Heartbleeding: The Rise of Wi-Fi Spoofing Attack Via
  Geolocation API}.
\newblock In {\em ACM Conference on Computer and Communications Security
  (CCS)}, 2022.

\bibitem{henderson2023space}
Stacey Henderson and Joel Lisk.
\newblock {Space War = Space Money? Are Commercial Actors the New Frontier for
  War}.
\newblock 2023.

\bibitem{isniff}
hubert3.
\newblock {iSniff GPS}, 2023.
\newblock \url{https://github.com/hubert3/iSniff-GPS/}.

\bibitem{hurley2004wardriving}
Chris Hurley.
\newblock {\em {WarDriving: Drive, Detect, Defend: A Guide to Wireless
  Security}}.
\newblock Elsevier, 2004.

\bibitem{oui}
IEEE.
\newblock {MAC Address Block Large (MA-L)}, 2023.
\newblock \url{https://standards-oui.ieee.org/oui/oui.txt}.

\bibitem{itzhak2023russian}
Aviv Itzhak and Ur~Fer.
\newblock {Russian-Ukraine Armed Conflict: Lessons Learned on the Digital
  Ecosystem}.
\newblock {\em {International Journal of Critical Infrastructure Protection}}.

\bibitem{starlinkinvasion}
Hyunjoo Jin.
\newblock {Musk says Starlink active in Ukraine as Russian invasion disrupts
  internet}.
\newblock {\em Reuters}, 2022.
\newblock
  \url{https://www.reuters.com/technology/musk-says-starlink-active-ukraine-russian-invasion-disrupts-internet-2022-02-27/}.

\bibitem{martin2017study}
Jeremy Martin, Travis Mayberry, Collin Donahue, Lucas Foppe, Lamont Brown,
  Chadwick Riggins, Erik~C Rye, and Dane Brown.
\newblock {A Study of MAC Address Randomization in Mobile Devices and When it
  Fails}.
\newblock {\em Privacy Enhancing Technologies Symposium (PETS)}, 2017.

\bibitem{matte2016defeating}
C{\'e}lestin Matte, Mathieu Cunche, Franck Rousseau, and Mathy Vanhoef.
\newblock {Defeating MAC Address Randomization through Timing Attacks}.
\newblock In {\em ACM Conference on Security and Privacy in Wireless and Mobile
  Networks (WiSec)}, 2016.

\bibitem{china-infrax}
Paul Mozur, Daisuke Wakabayashi, and Nick Wingfield.
\newblock {Apple Opening Data Center in China to Comply With Cybersecurity
  Law}, 2017.
\newblock
  \url{https://www.nytimes.com/2017/07/12/business/apple-china-data-center-cybersecurity.html}.

\bibitem{mylnikov}
Alexander Mylnikov.
\newblock {Geo-Location API Download Section}, 2024.
\newblock \url{https://www.mylnikov.org/download}.

\bibitem{openwifi}
openwifi.su.
\newblock {OpenWifi.su Dataset}, 2021.
\newblock \url{http://openwifi.su/db/}.

\bibitem{openbmap}
radiocells.org.
\newblock {OpenBMap Dataset}, 2021.
\newblock \url{https://radiocells.org/}.

\bibitem{ray2023starlink}
Kaushik Ray and William Selvamurthy.
\newblock {Starlink's Role in Ukraine}.
\newblock {\em Journal of Defence Studies}, 2023.

\bibitem{police-auctions-oakland}
Richard Roberts, Julio Poveda, Raley Roberts, and Dave Levin.
\newblock {Blue Is the New Black (Market): Privacy Leaks and Re-Victimization
  from Police-Auctioned Cellphones}.
\newblock In {\em IEEE Symposium on Security and Privacy}, 2023.

\bibitem{ipvseeyou}
Erik~C Rye and Robert Beverly.
\newblock {IPvSeeYou: Exploiting Leaked Identifiers in IPv6 for Street-Level
  Geolocation}.
\newblock In {\em IEEE Symposium on Security and Privacy}, 2023.

\bibitem{stopskyhook}
MG~Siegler.
\newblock {In April, Apple Ditched Google And Skyhook In Favor Of Its Own
  Location Databases }.
\newblock {\em Tech Crunch}, 2010.
\newblock \url{https://techcrunch.com/2010/07/29/apple-location/}.

\bibitem{skyhook}
Skyhook.
\newblock {Skyhook Wi-Fi Location}, 2023.
\newblock \url{https://www.skyhook.com/wifi-location-solutions}.

\bibitem{tippenhauer2009attacks}
Nils~Ole Tippenhauer, Kasper~Bonne Rasmussen, Christina P{\"o}pper, and Srdjan
  {\v{C}}apkun.
\newblock {Attacks on Public WLAN-Based Positioning systems}.
\newblock In {\em ACM Conference on Mobile Systems, Applications, and Services
  (MobiSys)}, 2009.

\bibitem{vanhoef2016mac}
Mathy Vanhoef, C{\'e}lestin Matte, Mathieu Cunche, Leonardo~S. Cardoso, and
  Frank Piessens.
\newblock {Why MAC Address Randomization is not Enough: An Analysis of Wi-Fi
  Network Discovery Mechanisms}.
\newblock In {\em Asia Conference on Computer and Communications Security (ASIA
  CCS)}, 2016.

\bibitem{wigle}
WiGLE.
\newblock {WiGLE -- All the Networks. Found by Everyone.}, 2023.
\newblock \url{https://wigle.net}.

\end{thebibliography}

\begin{acronym}
    \acro{AP}{access point}
    \acro{AKM}{Authentication and Key Management}
    \acro{BSSID}{Basic Service Set Identifier}
    \acro{ESS}{Extended Service Set}
    \acro{CPE}{Customer Premises Equipment}
    \acro{EAP}{Extensible Authentication Protocol}
    \acro{ISP}{Internet Service Provider}
    \acro{PSK}{Pre-Shared Key}
    \acro{LAN}{Local Area Network}
    \acro{MAC}{Media Access Control}
    \acro{OUI}{Organizationally Unique Identifier}
    \acro{WLAN}{Wireless Local Area Network}
    \acro{WPS}{Wi-Fi-based Positioning System}
\end{acronym}

\end{document}